\documentclass[10pt,newlfont,aps,pra,superscriptaddress,twocolumn]{revtex4}
\usepackage{color}
\usepackage{float}
\def\hs{\hspace{0.5cm}}
\def\ea{{\it et al.}}
\def\ft{\hspace{0.1cm}}
\def\aho{$a_{ho}\,=\,\sqrt{\hbar/(2 m\omega_{ho})}$}
\def\hw{$\hbar\omega_{ho}$}
\def\Rhw{$\hbar\omega_{ho}/\sqrt{\Gamma}$}
\def\hs{\hspace{0.5cm}}
\usepackage[pdftex]{graphicx}
\usepackage{sidecap}
\usepackage{atbeginend}
\begin{document}

\title{Conditions for order and chaos in the dynamics of a trapped Bose--Einstein condensate in coordinate and energy space}
\author{Roger R. Sakhel}
\affiliation{Department of Physics, Faculty of Science, Isra University, Amman 11622, Jordan}
\affiliation{The Abdus Salam International Center for Theoretical Physics, Strada Costiera 11, 34151 Trieste, Italy}
\author{Asaad R. Sakhel}
\affiliation{Department of Physics and Basic Sciences, Faculty of Engineering Technology, Al-Balqa Applied University, Amman 11134, Jordan}
\affiliation{The Abdus Salam International Center for Theoretical Physics, Strada Costiera 11, 34151 Trieste, Italy}
\author{Humam B. Ghassib}
\affiliation{Department of Physics, University of Jordan, Amman 11942, Jordan}
\author{Antun Balaz}
\affiliation{Scientific Computing Laboratory, Institute of Physics Belgrade, University of Belgrade, 
Pregrevica 118, 11080 Belgrade, Serbia}

\begin{abstract}
We investigate numerically conditions for order and chaos in the dynamics of an 
interacting Bose--Einstein condensate (BEC) confined by an external trap cut off by a hard-wall 
box potential. The BEC is stirred by a laser to induce excitations manifesting as irregular 
spatial and energy oscillations of the trapped cloud. Adding laser stirring to the external 
trap results in an effective time-varying trapping frequency in connection with the dynamically 
changing combined external+laser potential trap. The resulting dynamics are analyzed by plotting 
their trajectories in coordinate phase space and in energy space. The Lyapunov exponents are 
computed to confirm the existence of chaos in the latter space. Quantum effects and trap anharmonicity 
are demonstrated to generate chaos in energy space, thus confirming its presence and implicating 
either quantum effects or trap anharmonicity as its generator. The presence of chaos in energy space 
does not necessarily translate into chaos in coordinate space. In general, a dynamic trapping frequency 
is found to promote chaos in a trapped BEC. An apparent means to suppress chaos in a trapped BEC is 
achieved by increasing the characteristic scale of the external trap with respect to the condensate size.
\end{abstract}
\date{\today}
%\pacs{03.75.-b,67.85.De,67.85.Jk}
\maketitle

\section{Introduction}\label{sec:introduction}

\hs Although the literature on chaos is vast
\cite{Kodba:2005,Filho:2000,Gertjerenken:2010,Muruganandam:2002,Xiong:2010,Katz:2010,Chong:2004,Brezinova:2011,Tomsovic:1991,Diver:2014,Jaouadi:2010,Zhu:2009,Martin:2009,Horsely:2014,Coullet:2001,Kapulkin:2008,Zhang:2011,Liu:1997,Brambila:2013,Brandstaeter:1983,Gardiner:2002,Cheng:2010,Hai:2008,Brezinova:2012}, 
there is still quite a number of phenomena that need to be explored and understood that would enable 
full control of chaos. Perhaps a deeper examination of the mechanisms causing chaos is timely. This 
is necessary, for example, in the construction of quantum computers 
\cite{Zhang:2011,Levi:2003,Georgeot:2000,Weitenberg:2011,Pachos:2003} because chaos severely reduces the 
accuracy of a quantum computational process and can even destroy it. The Bose--Einstein condensation (BEC) 
community is currently interested in eliminating chaos in the dynamics of a BEC to achieve highly accurate 
quantum computers of the future. There have also been investigations on the chaotic quantum billiard 
\cite{Katz:2010,Tomsovic:1991,Milner:2001,Zhang:2004,Wimberger:2014}, which is a dynamical system deeply 
related to, and which is used to explain some of the results of, the present study. In general, the present 
work seeks ways for controlling chaos by obtaining a deeper understanding of the mechanisms that promote it. 
What is also particular about the present work is that it explicitly examines chaos in energy space. This has 
rarely been done before to the best of our knowledge.

\hs The chaotic dynamics of a BEC has drawn substantial interest in the last decade or two. Theoretical 
studies have included chaotic oscillations in an attractive BEC \cite{Muruganandam:2002}, chaos in optical 
lattices \cite{Chacon:2008,Horsely:2014}, the depletion of a BEC in a time-dependent trap \cite{Castin:1997}, 
the Gross-Pitavskii equation (GPE) with a chaotic potential \cite{Gardiner:2000}, coherence and instabilitiy 
in a BEC driven by periodic $\delta-$kicks \cite{Billam:2012}, finite-temperature non-equlibrium dynamics in 
a BEC \cite{Billam:2013}, as well transitions to instability in a kicked BEC \cite{Zhang:2004,Liu:2006}. 
Experimentally, there have been studies of dynamical instabilities of BECs in optical lattices 
\cite{Fallani:2004,Ferris:2008}. In this paper, we provide a comprehensive investigation of aspects of chaotic 
dynamics present in a two dimensional GPE.

\hs The goals of the present work are: (1) to obtain conditions for order and chaos in the dynamics of an 
interacting trapped BEC; and (2) to confirm the existence of chaos in its energy space. Our major task is 
to understand the origins of chaos in a trapped BEC, particularly when the trapping potential is time-dependent. 
The existence of chaos is confirmed by well-established methods, such as the phase-space trajectories 
\cite{Muruganandam:2002}, the energy-space trajectories, and the Lyapunov exponent \cite{Wimberger:2014,Kodba:2005}.

\hs We consider a trapped BEC excited using a red- or blue-detuned laser potential (RDLP or BDLP, respectively). 
The dynamic stirring causes the overall trapping frequency to vary with time, destroys frequency locking, and 
thereby causing chaos \cite{Diver:2014}. In addition, the blue (red)-detuned laser tends to reduce (increase) 
the phase-space density \cite{Garrett:2011} available for excitations in the combined laser$+$external potential 
trap. Indeed, a comparison between the latter effects of different phase-space densities unmasks a considerable 
difference in the dynamics that is strongly related to the way a laser modifies the energy-level structure of 
the external trapping potential. This difference enables the effect of phase-space density on the occurrence of 
chaos to be analyzed. Moreover, the role of quantum effects \cite{Kapulkin:2008} and trap anharmonicity is particularly 
revealed in the generation of spatial and energy chaos. Order in the spatial dynamics is then demonstrated not 
to imply order in the energy dynamics. Moreover, the conclusions reached here discourage in particular the use 
of an anharmonic trap to surround an optical lattice, e.g., when it comes to the transport of atomic qubits using 
an optical tweezer to implement collisional quantum gates \cite{Weitenberg:2011}. This is because the anharmonic 
trap can cause chaotic oscillations inside the system, which can destroy the process of quantum computation.

\hs Although BECs excited by stirrers have been addressed both experimentally \cite{Stiessberger:2000,Fujimoto:2010,Fujimoto:2011,Jackson:2000,Radouani:2004,Caradoc:1999,Caradoc:2000,Neely:2010,Engels:2007,Onofrio:2000,Madison:2000,Raman:2001,Raman:1999,Madison:2001,Horng:2009} 
and theoretically \cite{Proukakis:2006,Diener:2002,Aioi:2011,Uncu:2008,Weitenberg:2011,Carpentier:2008}, using blue 
\cite{Neely:2010,Raman:1999,Onofrio:2000,Raman:2001} as well as red-detuned lasers 
\cite{Madison:2000,Madison:2001,Hammes:2002,Garrett:2011,Scherer:2007,Tuchendler:2008,Stamper-Kurn:1998,Comparat:2006,Jacob:2011,Gustavson:2002,Barrett:2001,Schulz:2007}, 
very little attention has been paid to energy dynamics such as the soliton energy \cite{Parker:2003,Parker:2010,Proukakis:2004} 
and the total energy \cite{Law:2000}. In addition, a detailed examination of the effects of different phase-space densities 
(different laser amplitudes) is still lacking. We therefore revisit our previous systems \cite{Sakhel:2011,Sakhel:2013} 
with the same excitation methods and analyze their dynamics from a different perspective.

\hs The organization of the paper is as follows. In Sec.~\ref{sec:method}, the system of the present study is 
introduced along with our motivation. Then the GPE with the stirring laser potential is briefly discussed. 
Next, the physical observables are presented with the Lyapunov exponent acting as an important measure for the degree 
of chaos. A mode expansion of the GPE wavefunction is also considered from which further information about the chaos 
in the wavefunction is obtained. In addition, the units and numerics are outlined. In Sec.~\ref{sec:results}, the 
results of the simulations are displayed and discussed. The effect of the phase--space density, trapping frequency, 
trapping anharmonicity, and quantum effects on chaos are explored in a rigorous manner. In Sec.~\ref{sec:analysis-and-proof-of-chaos}, 
the results are analyzed. The irregular dynamics is rigorously tested for order and chaos by well-established methods. 
In Sec.~\ref{sec:validityGPE}, the validity of our GPE approach is established. The paper ends with conclusions in 
Sec.~\ref{sec:conclusions}. In Appendix \ref{app:effective-trapping-frequency}, equations are derived that explain the 
behavior of the effective trapping frequency of the laser$+$trap as a function of position and time.

\section{Method}\label{sec:method}

\hs The system is a trapped two-dimensional (2D) BEC cut off by a hard-wall box potential (BP) boundary 
\cite{Sakhel:2011,Sakhel:2013} and excited by a stirring laser. The external trap varies from harmonic to extremely 
anharmonic. The split-step Crank--Nicolson (CN) method \cite{Muruganandam:2009,Vudragovic:2012} was invoked to solve 
the 2D time-dependent Gross--Pitaevskii equation (TDGPE) in real time (see Fig.~2 of Ref.~\cite{Sakhel:2013}). The 
calculations were conducted using the computing cluster of the Max Planck Institute for Physics of Complex Systems, 
Dresden, Germany. In essence, this was a heavy computational project where for times of order $t\sim 10^4$ several 
days of CPU time were required to complete simulations.

\subsection{Motivation}

\hs The prime motivation in exploring this system is to study chaos in low dimensions. The role played by the hard-wall 
boundaries is noteworthy as they generate complicated structures in the density patterns of a trapped BEC, including 
those from the nonlinear Talbot effect \cite{Sakhel:2011,Ruostekoski:2001}. These patterns arise from the self-interference 
of an expanding BEC with reflections coming in from the hard walls. Hard walls are realized experimentally by forming 
sheets of light \cite{Bongs:1999}. Energy is thereby contained, and can be used to excite the BEC to very high energies. 
Once excited, it remains in these states for times long enough for chaotic behavior to be explored. Our study has been 
impelled by a quite relevant investigation by Fujimoto and Tsubota \cite{Fujimoto:2011} who studied vortex nucleation 
in a harmonically trapped 2D BEC via an oscillating barrier; however they did not address chaos. Another incentive has 
been provided from a study of phase effects in a harmonically trapped BEC which is periodically driven to chaotic 
behavior \cite{Deng:2012}.

\hs The importance of the dimple potential (RDLP) is worth underlining and can be understood from the following points: 
(1) It increases the phase--space density of the trapped BEC by introducing a richer energy--level structure; (2) It 
is able to trap and split a fragment from a BEC and to transport it away; and (3) It is experimentally realizable and 
has been used in quite a number of works. Experimentally, Garrett \ea\ \cite{Garrett:2011} studied the formation of a 
BEC in a cigar trap to which a dimple potential was added. Jacob \ea\ \cite{Jacob:2011} produced in it a BEC of sodium 
atoms. Theoretically, it has been proposed \cite{Diener:2002} and used \cite{Gardiner:2002,Uncu:2008} to model the kinetics 
of BEC \cite{Dutta:2015} as well as in an analysis of a BEC in an optical cavity driven by an external beam \cite{Diver:2014}. 
The latter work demonstrated that the essential features of the chaotic behavior of a BEC are low-dimensional.

\subsection{Gross--Pitaevskii equation and laser potential}

\hs The 2D TDGPE, as stated in Refs.~\cite{Sakhel:2011,Sakhel:2013}, is

\begin{eqnarray}
&&\left[-\frac{\partial^2}{\partial x^2}\,-
\frac{\partial^2}{\partial y^2}\,+\,
\tilde{V}(x,y;t)\,+\,{\cal G}\left|\varphi(x,y;t)\right|^2\,-\,
\right.\nonumber\\
&&\left.i\frac{\partial}{\partial t}\right]\varphi(x,y;t)\,=\,0, 
\label{eq:TDGPE}
\end{eqnarray}

where

\begin{equation}
{\cal G}=\frac{4N a_{s}}{\ell}\sqrt{2\pi\lambda}, 
\label{eq:N-coupling-constant}
\end{equation}

is the coupling constant with $N$ the number of particles, $a_s$ the $s$-wave scattering length, 
$\ell=\sqrt{\hbar/m\omega_{ho}}$ a length scale, and $\lambda\,=\,\omega_z/\omega_{ho}$ an anisotropy parameter 
determining the width of the ground-state solution in the $z$-direction, $\phi_o(z)$ with $\omega_z$ the trapping 
frequency perpendicular to the plane of the BEC. As demonstrated in Ref.~\cite{Muruganandam:2009}, the $z$-dependence 
of the 3D TDGPE is integrated out to obtain the 2D form Eq.~(\ref{eq:TDGPE}). $\varphi(x,y;t)$ is the wavefunction 
of the system, where $\int_{-\infty}^{+\infty} dx \int_{-\infty}^{+\infty} dy |\varphi(x,y;t)|^2=1$. As before, 
$\tilde{V}(x,y;t)\,=\,V(x,y;t)/\hbar\omega_{ho}$ is an external potential including the stirring laser and is given by

\begin{eqnarray}
&&\tilde{V}(x,y;t)\,=\,\nonumber\\
&&\frac{\sigma}{4}\left(|x|^{p_1}\,+\kappa\,|y|^{p_2}\right)\,+\,
A \exp\{-\beta[x^2\,+\,(y-vt)^2]\}.\nonumber\\
\label{eq:combined-potential}
\end{eqnarray}

Here $\omega_{ho}$ is the trapping frequency, $\sigma$ the strength of the external potential with exponents 
$p_1$ and $p_2$, $\kappa$ the anisotropy parameter, $A$ the amplitude of the stirrer ($A>0$ for BDLP and $A<0$ 
for RDLP), $\beta$ the exponent determining the width of the stirrer, and $v$ its velocity. In both cases, the 
stirrer sweeps the BEC starting from the center of the system at time $t$=0 and moving towards the hard wall in 
the $+y$-direction. The stirrer exits the BP without returning. $\varphi(x,y;t)$ as well as its gradient 
$\nabla\varphi(x,y;t)$ are assumed to be zero at the BP-boundary. This is to enforce the hard-wall effect, i.e., 
imposing a potential of infinite height.

\subsection{Energy components and chaos}

\hs The total energy is given by \cite{Dalfovo:1999,Sakhel:2013}

\begin{eqnarray}
&&E(t)\,=\,\int d^2\mathbf{r} \left[|\nabla \varphi(x,y;t)|^2\,+
\,\tilde{V}(x,y;t)|\varphi(x,y;t)|^2\,+\,\right.\nonumber\\
&&\left.\frac{{\cal G}}{2}|\varphi(x,y;t)|^4\right],
\label{eq:GP-Energy-Functional}
\end{eqnarray}

where the limits of the integration are only over the area of the BP. According to Ref.~\cite{Pethick:2002}, 
Eq.~(\ref{eq:GP-Energy-Functional}) is separated into four terms
\begin{equation}
E(t)=E_{zp}(t)+E_{flow}(t)+E_{osc}(t)+E_{int}(t),
\label{eq:Energy-Components}
\end{equation}

keeping in mind that, after the stirrer is removed from the BP, $E(t)$ remains constant while its various 
contributions still vary with time. Putting the wavefunction in polar form, $\varphi(x,y;t)\,=\,\sqrt{\rho(x,y;t)}\exp[i\phi(x,y;t)]$ 
with $\rho(x,y;t)=|\varphi(x,y;t)|^2$ the density and $\phi(x,y;t)$ the phase, the zero-point kinetic energy becomes

\begin{equation}
E_{zp}(t)=\int d^2\mathbf{r}\left[\nabla \sqrt{\rho(x,y;t)}\right]^2,
\label{eq:Ezp}
\end{equation}

the kinetic energy of particle flow

\begin{equation}
E_{flow}(t)=\int d^2\mathbf{r}(\nabla \phi)^2 \rho(x,y;t),
\label{eq:Eflow}
\end{equation}

the combined trap potential energy

\begin{equation}
E_{osc}(t)=\int d^2\mathbf{r}\tilde{V}(x,y;t)\rho(x,y;t),
\label{eq:Eosc}
\end{equation}

and finally the interaction energy

\begin{equation}
E_{int}(t)=\frac{{\cal G}}{2}\int d^2\mathbf{r}\rho(x,y;t)^2.
\end{equation}

Note that the total kinetic energy is given by

\begin{equation}
E_{kin}(t)=E_{zp}(t)+E_{flow}(t).
\label{eq:Ekin}
\end{equation}

\hs Comparisons can be made between the dynamics of each of these energy terms for two phase-space densities, 
obtained by applying two stirring lasers: a barrier with $A>0$ and a well with $A<0$. Chaos in energy space 
manifests itself by any irregular oscillations in the energy components and by plotting the trajectories $(E,\dot{E})$, 
where $E$ denotes the specific component and $\dot{E}=dE/dt$ the time derivative.

\subsection{Radial oscillations and chaos}

\hs In coordinate space, the root-mean-square (RMS) radius $R_{rms}=\sqrt{\langle r(t)^2\rangle}$ of the trapped 
cloud is computed using \cite{Sakhel:2013}.

\begin{equation}
\sqrt{\langle r(t)^2\rangle}\,=\,
\left[\int |\varphi(x,y;t)|^2 (x^2+y^2)\,d^2\mathbf{r}\right]^{1/2}.
\label{eq:rms-radius}
\end{equation}

Chaos is signaled by the irregular oscillatory behavior of $R_{rms}$ and by plotting the trajectories $(X,\dot{X})$ 
in phase space, where $X=\sqrt{\langle r(t)^2\rangle}$ and $\dot{X}=d\sqrt{\langle r(t)^2\rangle}/dt$.

\subsection{Lyapunov exponent}

\hs Another very reliable test for chaos is the Lyapunov exponent ${\cal L}$ \cite{Kodba:2005,Wimberger:2014}, 
which provides a quantitative measure. If after a very long simulation time ${\cal L}$ remains positive and almost 
constant, then this is a strong indication of chaotic dynamics. When ${\cal L}$ is zero or negative, then chaos is 
absent. To calculate ${\cal L}$, a nonlinear time-series analysis of the various observables is implemented using the 
package of Kodba \ea\ \cite{Kodba:2005}. ${\cal L}$ is calculated using the expression \cite{Kodba:2005}

\begin{equation}
{\cal L}=\frac{1}{M t_{evolv}}\sum_{\ell=1}^{M}{\rm ln}
\left(\frac{L_{evolv}^{(\ell)}}{L_0^{(\ell)}}\right),
\label{LyapEquation}
\end{equation}

where $L_0$ is the Euclidean distance between an initial point $\mathbf{p}(0)$ in the embedding space and its 
nearest neighbor, $\tau$ is the embedding delay, and $L_{evolv}$ is the final distance between them after an 
evolution for a time-step $t_{evolv}$. After each $t_{evolv}$, a replacement step $\ell$ is attempted in which 
the code looks for a new nearest neighbor of the evolved initial point. A number $M$ of replacement steps is 
attempted. The point $\mathbf{p}$ is defined by the vector sequence

\begin{equation}
\mathbf{p}(i)=(x_i,x_{i+\tau},x_{i+2\tau},...,x_{i+(m-1)\tau}),
\label{eq:vectorP}
\end{equation}

obtained from the time series, where $m$ is the embedding dimension and $i$ the time. The variable $x$ stands 
for any observable.

\subsection{Mode expansion}

\hs Next, the solution to the TDGPE, $\varphi(x,y;t)$, is expanded into different sets of states: the harmonic 
oscillator (HO) function ${H_n} (u)$, the Legendre polynomials ${P_n}(u)$, and the Cosine function $\cos(n\pi u)$ 
with $u$ $\in$ $(x,y)$ and $n\in\{n_x,n_y\}$ an integer. For the HO case, the wavefunction becomes the double sum

\begin{eqnarray}
&&\varphi(x,y;t)\,=\,\nonumber\\
&&\sum_{n_x=0} \sum_{n_y=0} C_{n_x,n_y}(t) B_{n_x} H_{n_x}(x) B_{n_y}
H_{n_y}(y) e^{-(x^2+y^2)/2}, \nonumber\\
\label{eq:wave-function-expansion-into-HO-states}
\end{eqnarray}

$n_x$ and $n_y$ being the HO quantum numbers, $B_{n}=(n!\,2^{n} \sqrt{\pi})^{-1/2}$ the normalization constant 
of $H_n(x)\exp(-x^2/2)$, and $C_{n_x,n_y}(t)$ are time-dependent mode amplitudes that describe the population 
dynamics of the states $(n_x,n_y)$. The $C_{n_x,n_y}(t)$ are obtained from Eq.~(\ref{eq:wave-function-expansion-into-HO-states}) 
using the orthogonality of the Hermite polynomials such that

\begin{eqnarray}
C_{n_x,n_y}(t)\,&=&\,\int_{-\infty}^{+\infty} dx \int_{-\infty}^{+\infty} dy 
\cdot\varphi(x,y;t)\times \; \nonumber\\ 
&&B_{n_x} H_{n_x}(x) B_{n_y} H_{n_y}(y) e^{-(x^2+y^2)/2}.
\label{eq:Cnxny}
\end{eqnarray}

For ${P_n} (u)$, one should rescale $u$ by the length of the system $L$ and use $B_n=\sqrt{(2n+1)/2}$ for the 
normalization constants; similarly for $\cos(n\pi u)$ where $B_n=1/\sqrt{L}$. In particular, the evolutionary 
patterns of $C_{n_x,n_y}(t)$ signal chaos or order in the population dynamics of the various basis states after 
an evaluation of their Lyapunov exponents. In addition, they are an important indicator of allowed and forbidden 
transitions between the HO states. In passing, it is noted that the expansion (\ref{eq:wave-function-expansion-into-HO-states}) 
is the same as that of the classical field approach \cite{Blakie:2008}, except that the mode amplitudes are extracted 
from a numerical solution of the TDGPE at $T=0$ K.

\subsection{Units}\label{sec:units}

\hs The units and numerics are the same as in our previous work \cite{Sakhel:2011,Sakhel:2013}; they are reviewed 
here briefly for reference purposes. ${\cal G}$, $A$, $v$, $\beta$, and $t$, all have the same units as before 
\cite{Sakhel:2011,Sakhel:2013}: The lengths and energies are in units of the trap \aho\ and \hw, respectively. $A$ is 
in units of \hw, $\beta$ in $(a_{ho})^{-2}$, $v$ in $a_{ho}$, $t = \tau\omega_{ho}$ is unitless, ${\cal G}$ is in 
$(\sqrt{2} a_{ho}^2)^{-1}$, and $\kappa$ is unitless. The energies Eqs.(\ref{eq:Energy-Components})-(\ref{eq:Ekin}) 
are in units of \hw.

\subsection{Numerics}\label{sec:numerics}

\hs Throughout, the following parameter settings are used. For the stirrer, we set $v=2$ and $\beta$=4, whereas 
$|A|$ ranges from 0 to 40. For the external trap, we set $\kappa$=1 such that it is always isotropic. Initially 
we set $p_1=p_2=2$ for a harmonic trap, but later set $p_1=p_2>2$ to explore the effect of trap anharmonicity. The 
number of particles for a given ${\cal G}$ can be evaluated from Eq.~(\ref{eq:N-coupling-constant}) with the following 
information. For $^{87}$Rb, the scattering length is $a_s=5.4$ nm and a suitable trapping frequency is 
$\omega_{ho}=2\pi\,\times\,25$ Hz \cite{Ruostekoski:2001}. The trap length is then $\ell=2.16$ ${\rm\mu}$m (ours 
is $a_{ho}=\ell/\sqrt{2}$). The anisotropy $\lambda$ used in Eq.~(\ref{eq:N-coupling-constant}) is set to 10 so 
that the width of the ground state $\phi_{o}(z)$ becomes extremely small along the $z-$direction and the system 
can be considered 2D. For this $\lambda$ and a value like ${\cal G}=10$ used here, the number of particles is 
$N\sim 117$ and the BEC is in the weakly interacting regime. The velocity $v$ by which the stirrer is moved can 
be converted to standard units by $v\,\rightarrow\,v a_{ho}\omega_{ho}$. Using the previous information, $v$=1 
in trap units is then equal to $2.4\times 10^{-4}$m/s. The BP length is $L=20$ ($a_{ho}$) {\it i.e.}, 
$-10\le x \le 10$ and $-10\le y \le 10$; that is in SI units the density becomes $n\sim N/L^2=1.254\times 10^{11}\hbox{m}^{-2}$ 
yielding $n a_s^2 \sim 10^{-6}$. The dynamics were mostly displayed for times up to $t$ = 20 corresponding to 
0.127~s, which is within experimental observation times, e.g., of Donley {\it et al.} \cite{Donley:2001}. In 
calculating Lyapunov exponents, the simulations were conducted for extended periods of $t=10000$ so as to 
positively confirm the presence of chaos in the observables [Eqs.~(\ref{eq:Ezp})-(\ref{eq:rms-radius})]. The 
simulations are initialized using Method (a) in Refs.~\cite{Sakhel:2011,Sakhel:2013}. As before, the results 
presented are in the transient stage of the simulation, {\it i.e.}, after, and not including, the initialization process.

\begin{SCfigure*}
\centering
\includegraphics[width=11.5cm,viewport=24 169 526 770,clip]{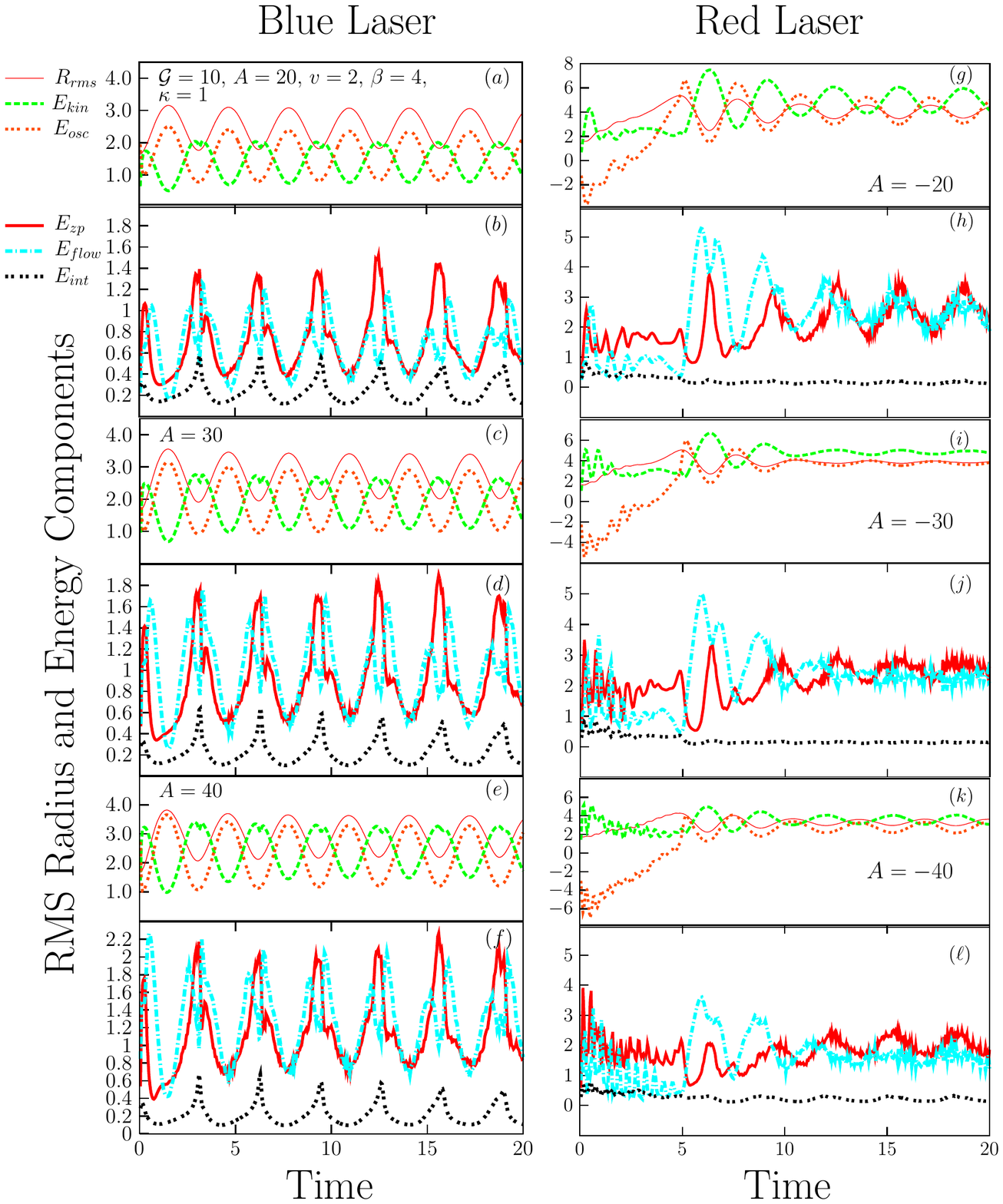}
\caption{(Color online) Spatial and energy dynamics of an interacting Bose gas confined by a two-dimensional 
harmonic trap cut off by a BP \cite{Sakhel:2011,Sakhel:2013} and various stirring laser amplitudes $A$. The side 
length of the BP is $L=20$ such that $x\in[-10,10]$ and $y\in[-10,10]$. The Bose gas is excited by a stirring 
blue-detuned laser ($A>0$ left column) and a red-detuned laser ($A<0$ right column). The parameters of the system 
are \cite{Sakhel:2011,Sakhel:2013}: ${\cal G} = 10$, $\beta = 4$, $v = 2$, $p_1=p_2=2$, and $\kappa = 1$. $A=20$ 
[frames ($a$) and ($b$)], 30 [($c$) and ($d$)], and 40 [($e$) and ($f$)]. $A=-20$ [($g$) and ($h$)], $-30$ [($i$) 
and ($j$)], and $-40$ [($k$) and ($\ell$)]. Solid line: $R_{rms}$; dashed line: $E_{kin}$; triple-dotted line: 
$E_{osc}$; thick solid line: $E_{zp}$; dashed-dotted line: $E_{flow}$; double-dotted line: $E_{int}$. $A$ is in 
units of \hw, $\beta$ in $(a_{ho})^{-2}$, $v$ in $a_{ho}$, ${\cal G}$ is in $(\sqrt{2} a_{ho}^2)^{-1}$, $R_{rms}$ 
in $a_{ho}$, and energy is in \hw. $\kappa$ and $t = \tau\omega_{ho}$ are unitless.}
\label{fig:ComponentsplotEvstG10B4T100Step5percentKseveralA}
\end{SCfigure*}%Fig.1

\section{Results and discussion}\label{sec:results}

\subsection{Effect of phase-space density (laser intensities)}

\hs Figure\ft\ref{fig:ComponentsplotEvstG10B4T100Step5percentKseveralA} demonstrates the dynamics of the 
energy components and radial size described by Eqs.~(\ref{eq:Ezp}$-$\ref{eq:rms-radius}) under the effect 
of a BDLP (left column) and an RDLP (right column) for various values of $A$ but fixed interactions ${\cal G}$ 
in a harmonic trap.

\subsubsection{Reduced phase-space density (blue-detuned laser)}

\hs Stirring using a BDLP reduces the phase space density of the trapped BEC yielding a small irregularity 
in the dynamics of its energy components and average radial size of its cloud. $R_{rms}$, $E_{kin}$, and 
$E_{osc}$ display ordered, sinusoidal, oscillatory patterns \cite{gpe-integrability}, whereas the dynamics 
of $E_{zp}$, $E_{flow}$, and $E_{int}$ are ---from an initial shrewd guess---apparently ordered, but not 
without some noise and irregularities. The oscillations in $E_{zp}$ arise from oscillations in the density 
$\rho(x,y;t)$, whereas those in $E_{flow}$ from oscillations in $\rho(x,y;t)$ and the phase $\phi(x,y;t)$. 
In addition, the oscillations in $\rho(x,y;t)$ arise from the center-of-mass oscillations of the trapped BEC 
cloud and continue with the same pattern even after the BDLP leaves the BP. The reason for this is explained 
in Sec.\ft\ref{sec:zero-point-motion-eff-trap-freq} below.

\hs A change in the amplitude of the BDLP yields no qualitative changes in the dynamics. This is because 
the BDLP introduces only a circular `hole' in the BEC with no introduction of additional lower energy levels, 
unlike the RDLP case, which increases the phase-space density. In essence, the effect of a blue-detuned laser 
is similar to a hard-disk-like obstacle which moves through a 2D fluid.

\subsubsection{Enlarged phase-space density (red-detuned laser)}

\hs RDLP stirring introduces a larger phase-space density that leads to higher degrees of irregular behavior 
and causes a strong time-dependent asymmetry of $|\varphi(x,y;t)|^2$ about the $x$-axis. This yields irregular 
interference patterns that translate to chaotic behavior. In comparison, the BDLP results in a much weaker 
asymmetry of $|\varphi(x,y;t)|^2$. Now, although the excitations induced by the RDLP cause somewhat regular 
oscillations in $R_{rms}$, $E_{kin}$, $E_{osc}$, and $E_{int}$, the other quantities like $E_{zp}$ and 
$E_{flow}$ reveal irregular oscillations throughout. For $E_{zp}$ and $E_{flow}$ this can be understood 
based on them being connected to gradients $\nabla|\varphi(x,y;t)|$ and $\nabla\phi(x,y;t)$, respectively, as 
they strongly change with time and with a high degree of randomness.

\hs The BEC fragment trapped inside the RDLP undergoes density oscillations at a frequency equivalent to 
the effective trapping frequency, $\omega_q$, of the combined trap that in turn yields oscillations in 
$\rho(x,y;t)$ as well. According to Eq.~(\ref{eq:omegax0vttRDLP}), $\omega_q$ rises with ``increasing" 
depth $A<0$ and because $E_{zp}$ and $E_{flow}$ are connected to the $\rho(x,y;t)$ via Eqs.~(\ref{eq:Ezp}) 
and (\ref{eq:Eflow}), their oscillation frequency at $t<5$ rises significantly. These oscillations are not 
observed in the BDLP case because $\omega_q$ remains roughly equivalent to that of the external trap and 
is smaller than that for the RDLP case. Also, the availability of energy levels and the dynamic $\omega_q$ 
can be argued to induce irregular oscillatory patterns because of the absence of frequency locking 
\cite{Diver:2014,Quasiperiodicity-chaos-transition}. Note that the influence of these initial strong 
excitations via the red-detuned laser remains even after it has left the trap, as can be seen by the irregular 
dynamics of $E_{zp}$ and $E_{flow}$ at $t>5$. The conclusion is that once chaos has been established, it is 
irreversible.

\subsection{Effective trapping frequency}\label{sec:zero-point-motion-eff-trap-freq}

\hs Earlier it was argued that irregularity (chaos according to Ref.~\cite{Kapulkin:2008}) is induced by 
quantum effects arising from the zero-point motion. The latter is controlled by $\omega_q$ and because a 
time-dependent trapping potential yields a dynamic $\omega_q$ (Appendix A) examining its role in irregular 
behavior as governed by different phase-space densities is therefore important. Now, $\tilde{V}(x,y;t)$ 
[Eq.~(\ref{eq:combined-potential})] has an effective $\omega_q$ defined as

\begin{equation}
\omega_q(x,y;t)=\sqrt{\frac{\partial^2\tilde{V}(x,y;t)}{\partial q^2}},
\label{eq:omegaq_effective_trapping_frequency}
\end{equation}

close to the minimum of $\tilde{V}(x,y;t)$ where $q\equiv x,y$, or $z$. That is, $\omega_q$ can vary 
spatially as well. It is however expected that $\tilde{V}(x,y;t)$ influences the patterns of the induced 
irregularities only when $\omega_q$ changes with time. Indeed, Eq.~(\ref{eq:omegaq_effective_trapping_frequency}) 
is only applicable to the red-detuned laser with $A<0$ and it cannot be applied to the blue-detuned 
laser with $A>0$, except when the BDLP is exactly centered at the origin of the external harmonic trap. 
At this point only, the BDLP is surrounded by a circular trough containing a circle of the potential minima 
loci. In that sense, the BDLP is only an obstacle and it turns out that its motion inside the BEC yields 
absolutely little change in $\omega_q$ in contrast to the RDLP. We provide further support for this conjecture 
in the following argument. In Fig.~\ref{fig:ComponentsplotEvstG10B4T100Step5percentKseveralA}, the blue laser 
induces oscillations in the dynamics which have almost the same pattern and frequency before and after the laser 
leaves the BP, whereas they are largely different for the red-detuned laser. One origin of this difference is 
that the blue-detuned laser does not trap any bosons unlike the red-detuned laser, and consequently the removal 
of the BDLP from the trap does not affect $\omega_q$. Therefore, the frequency of oscillations remain by and 
large controlled by the external harmonic trap. In contrast, the RDLP conveys a larger effective trapping 
frequency to the whole system. When the red-detuned laser leaves the trap, $\omega_q$ changes back to that 
of a pure harmonic oscillator. As a result, the dynamics display different behavior before and after the removal 
of the RDLP.

\begin{SCfigure*}
\centering
\includegraphics[width=11.5cm,viewport=71 138 577 763,clip]{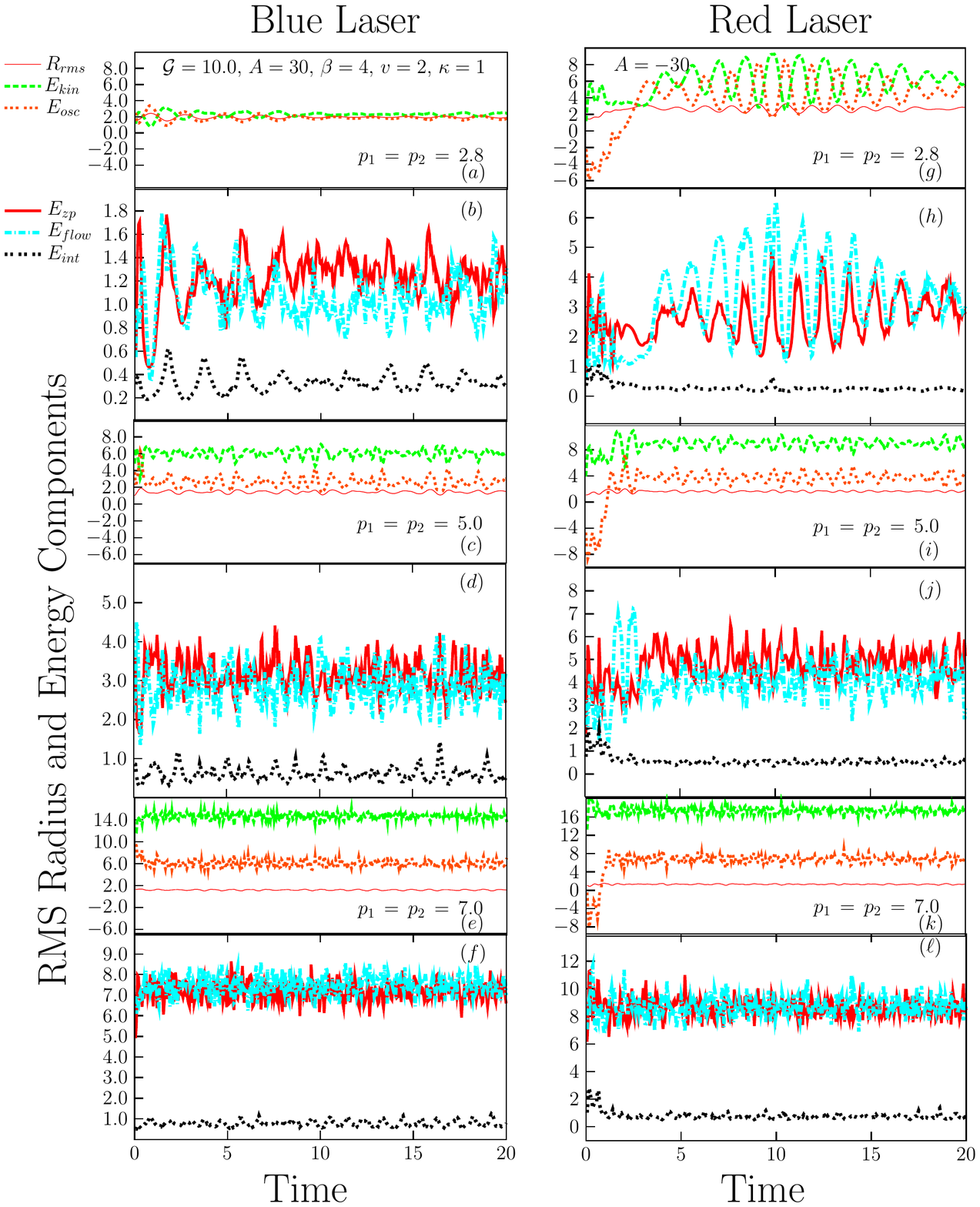}
\caption{(Color online) Same as in Fig.~\ref{fig:ComponentsplotEvstG10B4T100Step5percentKseveralA} with the same 
labels; but for varying anharmonicities $p_1=p_2$. The interactions are fixed at ${\cal G}=10$, the stirrer 
height is $A=30$ in the left column and its depth is $A=-30$ in the right column. Frames ($a$),($b$),($g$),($h$): 
$p_1=p_2=2.8$; ($c$),($d$),($i$),($j$): 5.0; ($e$), ($f$),($k$),($\ell$): 7.0. $A$ is in units of \hw, 
${\cal G}$ in $(\sqrt{2}a_{ho}^2)^{-1}$, $v$ in $a_{ho}$, $\beta$ in $(a_{ho})^{-2}$, and $t=\tau\omega_{ho}$ and 
$\kappa$ are unitless.}
\label{fig:ComponentsplotEvstG10B4T30Step5percentK1severalp}
\end{SCfigure*}%Fig.2

\subsection{Effect of trap anharmonicity}

\hs We now turn to anharmonic traps. From Fig.~\ref{fig:ComponentsplotEvstG10B4T30Step5percentK1severalp}, it 
is demonstrated that such traps mainly cause irregular oscillations in coordinate and energy space. Both sets of 
plots demonstrate an apparent increase in irregularity and frequency of the oscillations with increasing $p_1=p_2$. 
A similar irregularity was reported earlier by Mateos and Jos$\acute{e}$ \cite{Mateos:1998} for a particle inside 
a rigid BP with a periodically oscillating square-potential barrier inside the box. The latter behavior has been 
identified as chaotic purely from observations of energy oscillations.

\hs An increase in $p_1=p_2$ yields a larger $\omega_q$ [see Eqs.~(\ref{eq:omega_x.dependence_on_p}) and 
(\ref{eq:omega_y.dependence_on_p})], that in turn increases the frequency at which the BEC fragment oscillates 
inside the RDLP. For example, according to Eq.~(\ref{eq:omega_y}) for $p_1=p_2=7$, $\omega_q$ increases with 
time, and with it the number of HO modes, until the moving RDLP leaves the trap. The stronger anharmonicity 
generates oscillations in the energy components to become more irregular than in 
Fig.~\ref{fig:ComponentsplotEvstG10B4T100Step5percentKseveralA} and to `wash out' the difference between the 
blue and red-detuned laser. At this point, the BEC is in such a highly excited state with strong irregular 
dynamic behavior, that the vastly different phase-space densities can no longer be distinguished.

\begin{figure}[t!]
\includegraphics[width=8.0cm,viewport=163 366 439 764,clip]{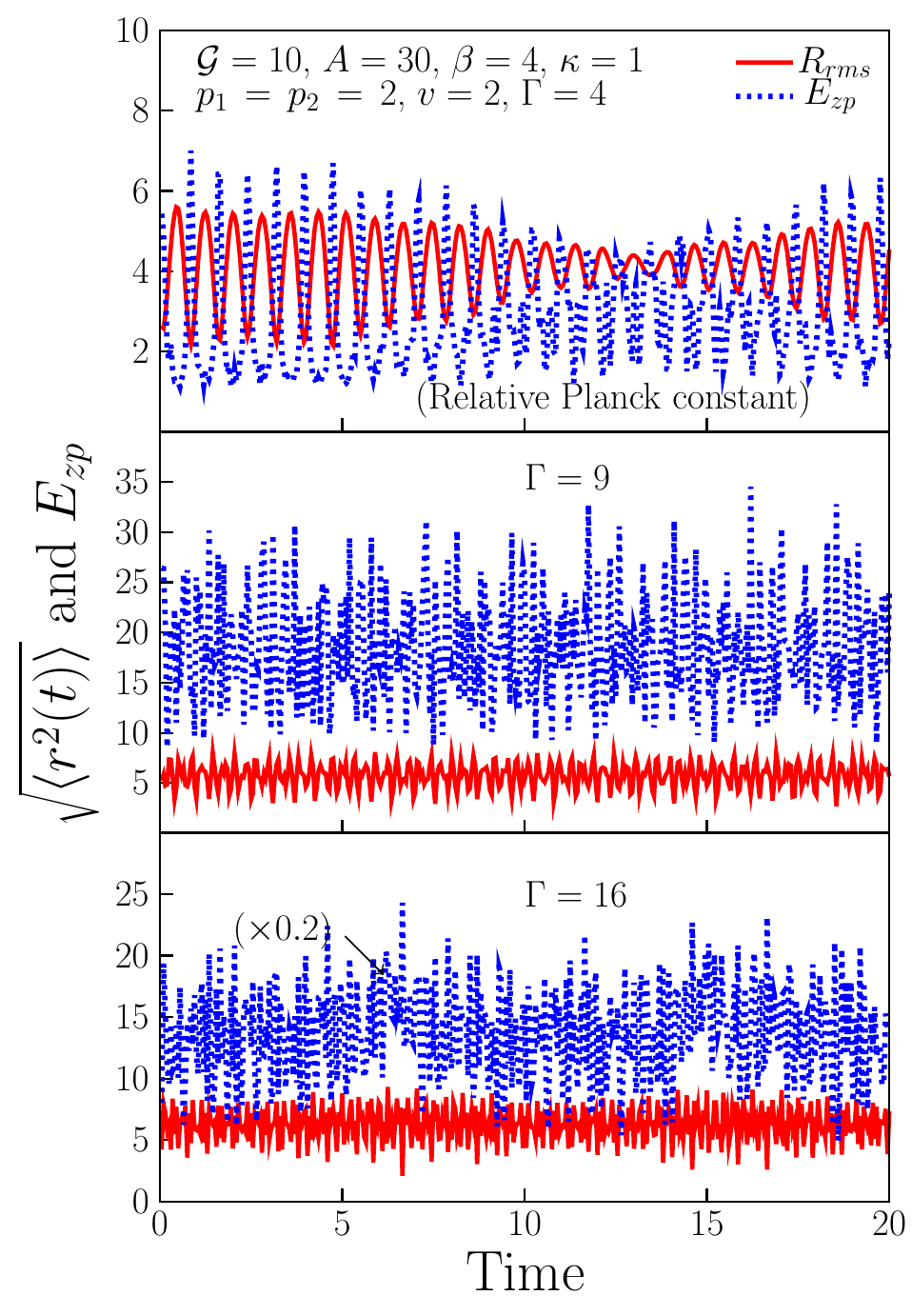}
\caption{(Color online) Quantum effects on the dynamics of a trapped Bose gas. The system is the same as in 
Fig.~\ref{fig:ComponentsplotEvstG10B4T100Step5percentKseveralA} for $A=+30$; but with increased relative 
constant $\Gamma$. The solid line is $R_{rms}(t)=\sqrt{\langle r^2(t)\rangle}$ and the dashed line $E_{zp}(t)$. 
Top frame: $\Gamma=4$; middle frame: $9$; and bottom frame: $16$. In the bottom frame $E_{zp}(t)$ is reduced 
by a multiplicative factor of $0.2$ to make $R_{rms}(t)$ visible and not to clutter the figure. $R_{rms}(t)$ 
is in units $a_{ho}$, $E_{zp}(t)$ in \Rhw\ (see text), $A$ in \Rhw, ${\cal G}$ in $(\sqrt{2}a_{ho}^2)^{-1}$, 
$v$ in $a_{ho}$, $\beta$ in $(a_{ho})^{-2}$, and $t=\tau\omega_{ho}/\sqrt{\Gamma}$ and $\kappa$ are unitless.}
\label{fig:QuantumeffectsRMSradiusvariousRelativehbarPropertiesBlueLaser}
\end{figure}%Fig.3

\begin{figure}[t!]
\includegraphics[width=8.0cm,viewport=58 399 309 702,clip]{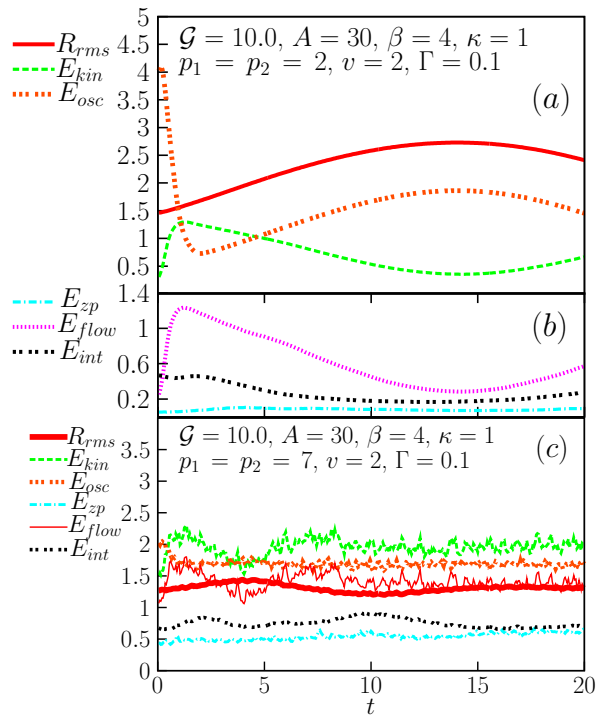}
\caption{(Color online) Frames ($a$) and ($b$): Dynamics of the harmonically trapped system in 
Fig.~\ref{fig:ComponentsplotEvstG10B4T100Step5percentKseveralA} at $A=+30$ for an artificially reduced 
relative constant $\Gamma=0.1$. Solid line: $R_{rms}(t)$; dashed line: $E_{kin}(t)$; triple-dotted line: 
$E_{osc}(t)$; dashed-dotted line: $E_{zp}(t)$; fine-dashed line: $E_{flow}(t)$; double-dotted line: $E_{int}(t)$. 
Frame ($c$) is as in frames ($a$) and ($b$); but for the anharmonically trapped system in 
Fig.~\ref{fig:ComponentsplotEvstG10B4T30Step5percentK1severalp} [$(e)$ and $(f)$] with $A=+30$. Thick solid 
line: $R_{rms}(t)$; dashed line: $E_{kin}(t)$; triple-dotted line: $E_{osc}(t)$; dashed-dotted line: $E_{zp}(t)$; 
thin solid line: $E_{flow}(t)$; and double-dotted line: $E_{int}(t)$. Lengths and energies are in units of 
the trap $a_{ho}$ and \Rhw, respectively. $A$ is in units of \Rhw, ${\cal G}$ in $(\sqrt{2}a_{ho}^2)^{-1}$, 
$v$ in $a_{ho}$, $\beta$ in $(a_{ho})^{-2}$, and $t=\tau\omega_{ho}/\sqrt{\Gamma}$ and $\kappa$ are unitless.}
\label{fig:QuantumeffectsRelativeHbarG10A30B4T20Beta0.1stack}
\end{figure}%Fig.4

\begin{figure}[t!]
\includegraphics[width=9.2cm,viewport = 178 455 589 702,clip]{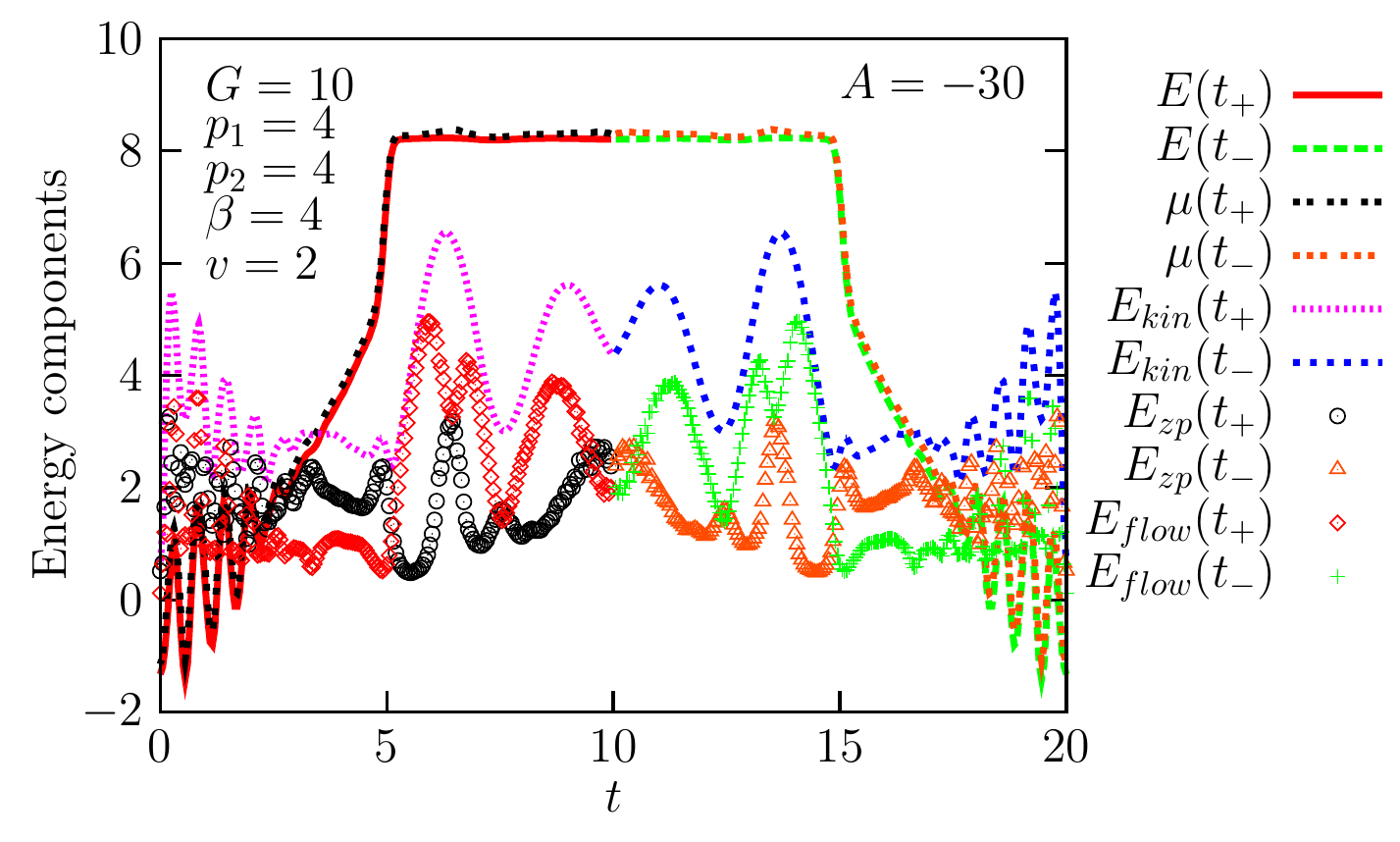}
\caption{(Color online) Checking for the presence of physical or numerical chaos in the Crank--Nicolson 
code applied in this work. The figure shows forward and backward evolving properties of a trapped BEC 
excited by an RDLP. The system is the same as in Fig.~\ref{fig:ComponentsplotEvstG10B4T100Step5percentKseveralA} 
for $A=-30$; except for $p_1=p_2=4$. The $\pm$ subscripts in the labels are for forward/backward evolution in 
time: Thick solid line: total energy $E(t_+)$; dashed line: $E(t_-)$; double-dotted line: chemical potential 
$\mu(t_+)$; triple-dotted line $\mu(t_-)$; fine-dotted line: kinetic energy $E_{kin}(t_+)$; dotted line: 
$E_{kin}(t_-)$; open circles: zero-point energy $E_{zp}(t_+)$; open triangles $E_{zp}(t_-)$; diamonds: 
kinetic flow energy $E_{flow}(t_+)$; and crosses: $E_{flow}(t_-)$. In this figure, the time at which time 
reversal begins is $t_0=10$ (see text); to the left of $t_0$, the properties are evolving forward in time $t$, 
whereas to the right of $t_0$, the time axis is for $t=2t_0-t^\prime$, where $t^\prime$ is decreasing as the 
properties are evolving backward in time. Lengths and energies are in units of the trap, $a_{ho}$ and \hw, 
respectively. $A$ is in units of \hw, ${\cal G}$ in $(\sqrt{2}a_{ho}^2)^{-1}$, $v$ in $a_{ho}$, $\beta$ in 
$(a_{ho})^{-2}$, and $t=\tau\omega_{ho}$ is unitless.}
\label{fig:plottimereversaltestG10A-30V2B4powx2powy2Energy}
\end{figure}%Fig5

\subsection{Quantum effects}\label{sec:zero-point}

\hs In this section, we follow Kapulkin and Pattanayak \cite{Kapulkin:2008} in an artificial modification 
of the relative Planck's constant to examine the role of quantum effects in the emergence of chaos or 
irregularity in the TDGPE. The latter is rescaled in a manner described below so as to introduce this 
constant via a multiplicative factor $\Gamma$ into the kinetic energy operator

\begin{equation}
\hat{H}_{kin}=-\Gamma\nabla^2,
\label{eq:hatH}
\end{equation}

from which it follows that $E_{zp}$ [Eq.~(\ref{eq:Ezp})] and $E_{flow}$ [Eq.~(\ref{eq:Eflow})] become 
weighted by $\Gamma$. One can justify the introduction of $\Gamma$ in Eq.~(\ref{eq:hatH}) based on the 
following argument. In Ref.~\cite{Kapulkin:2008}, a factor $\beta_0=\sqrt{\hbar/(m\ell_0^2 \omega_{ho})}$ 
was introduced that identifies a ratio between the characteristic scales associated with a BEC 
[$\ell=\sqrt{\hbar/(m\omega_{ho})}$] and the trap ($\ell_0$). Let us define $\ell_0=\sqrt{\hbar/(m\omega_0)}$, 
which can be different from $\ell$. Further, let us introduce an artificial Planck's constant $\hbar_0$ 
so that $\ell_0\,=\,\sqrt{\hbar/(m\omega_0)}\,\equiv\,\sqrt{\hbar_0/(m\omega_{ho})}$ and therefore

\begin{equation}
\sqrt{\Gamma}\,=\,\hbar/\hbar_0\,=\,\omega_0/\omega_{ho}\,=\,(\ell/\ell_0)^2.
\label{eq:gamma-explanation}
\end{equation}

That is, a change of the relative Planck's constant $\sqrt{\Gamma}$ amounts to a change in the ``relative" 
characteristic scale $\ell/\ell_0$. For brevity, $\Gamma=(\hbar/\hbar_0)^2$ is referred to as the 
relative constant and one can then consider rescaling the TDGPE

\begin{equation}
\left[-\frac{\hbar^2}{2m}\nabla^2\,+\,V(x,y;t)\,+\,g|\varphi|^2\right]\varphi\,=\
i\hbar\frac{\partial}{\partial t} \varphi.
\label{eq:TDGPE-SI-units}
\end{equation}

by a rescaling of the coordinates $x$ and $y$ in units of $a_{ho}\,=\,\ell_0/\sqrt{2}$. That is, one 
considers $\widetilde{x}\,=\,x/a_{ho}$ and $\widetilde{y}\,=\,y/a_{ho}$ and divides Eq.~(\ref{eq:TDGPE-SI-units}) 
by $\hbar_0\omega_{ho}$ so that it becomes

\begin{equation}
\left[-\Gamma\widetilde{\nabla}^2\,+\,\frac{1}{4}\widetilde{\sigma}\left(|\widetilde{x}|^n\,+\,|\widetilde{y}|^n)\,
+\,\widetilde{g}|\varphi|^2\right)\right]\varphi\,=\,i\frac{\partial}{\partial \widetilde{t}}\varphi,
\end{equation}

where $\widetilde{\sigma}\,=\,\sigma a_{ho}^n/(\hbar_0\omega_{ho})$, $\widetilde{g}\,=\,g/(\hbar_0\omega_{ho})$, 
and $\widetilde{t}\,=\,\omega_{ho} t/\sqrt{\Gamma}$. When $\Gamma$ is small, the characteristic scale associated with the 
trapping potential is large with respect to the condensate size. Consequently, as $\Gamma\rightarrow 0$ 
the BEC wavefunction tends to become localized rather like a wave packet representing a classical particle. 
When $\Gamma$ is large, the BEC extends over a considerable region and quantum effects become visible.

\hs By artificially increasing $\Gamma$ to values larger than 1, the frequencies of the spatial and 
energy oscillations increase and their dynamics become more irregular 
(Fig.~\ref{fig:QuantumeffectsRMSradiusvariousRelativehbarPropertiesBlueLaser}). This is consistent with 
results of Kapulkin and Pattanayak \cite{Kapulkin:2008}.

\hs The question then arises: will chaos vanish if we set $\Gamma$ below 1? 
Fig.~\ref{fig:QuantumeffectsRelativeHbarG10A30B4T20Beta0.1stack} displays the dynamics for a significantly 
lowered $\Gamma\,=\,0.1$ and it is obvious that the frequency of the oscillations is largely reduced in 
frames ($a$) and ($b$), as compared with that for 
Fig.~\ref{fig:QuantumeffectsRMSradiusvariousRelativehbarPropertiesBlueLaser}. A significant result then is 
that the quantum effects are the sole reason for any high-frequency oscillations in the BEC dynamics.

\hs If we examine Fig.~\ref{fig:QuantumeffectsRelativeHbarG10A30B4T20Beta0.1stack}$(c)$, then as a 
result of lowering $\Gamma$, the oscillations are seen to reduce substantially in $R_{rms}$, $E_{zp}$ 
and $E_{int}$ as compared with those in Fig.~\ref{fig:ComponentsplotEvstG10B4T30Step5percentK1severalp} 
[$(e)$ and $(f)$], but they nevertheless remain active in $E_{osc}$, $E_{kin}$ and $E_{flow}$. The 
persistence of oscillations in an anharmonic trap in the presence of reduced quantum effects is a 
manifestation of the role of anharmonicity in inducing this behavior in conjunction with the chaotic
billiard effect. In frames $(a)$ and $(b)$, chaos has vanished, demonstrating that the reduction of 
$\Gamma$ is one way to suppress chaos.

\begin{SCfigure*}
\includegraphics[width=13cm,viewport=65 334 519 731,clip]{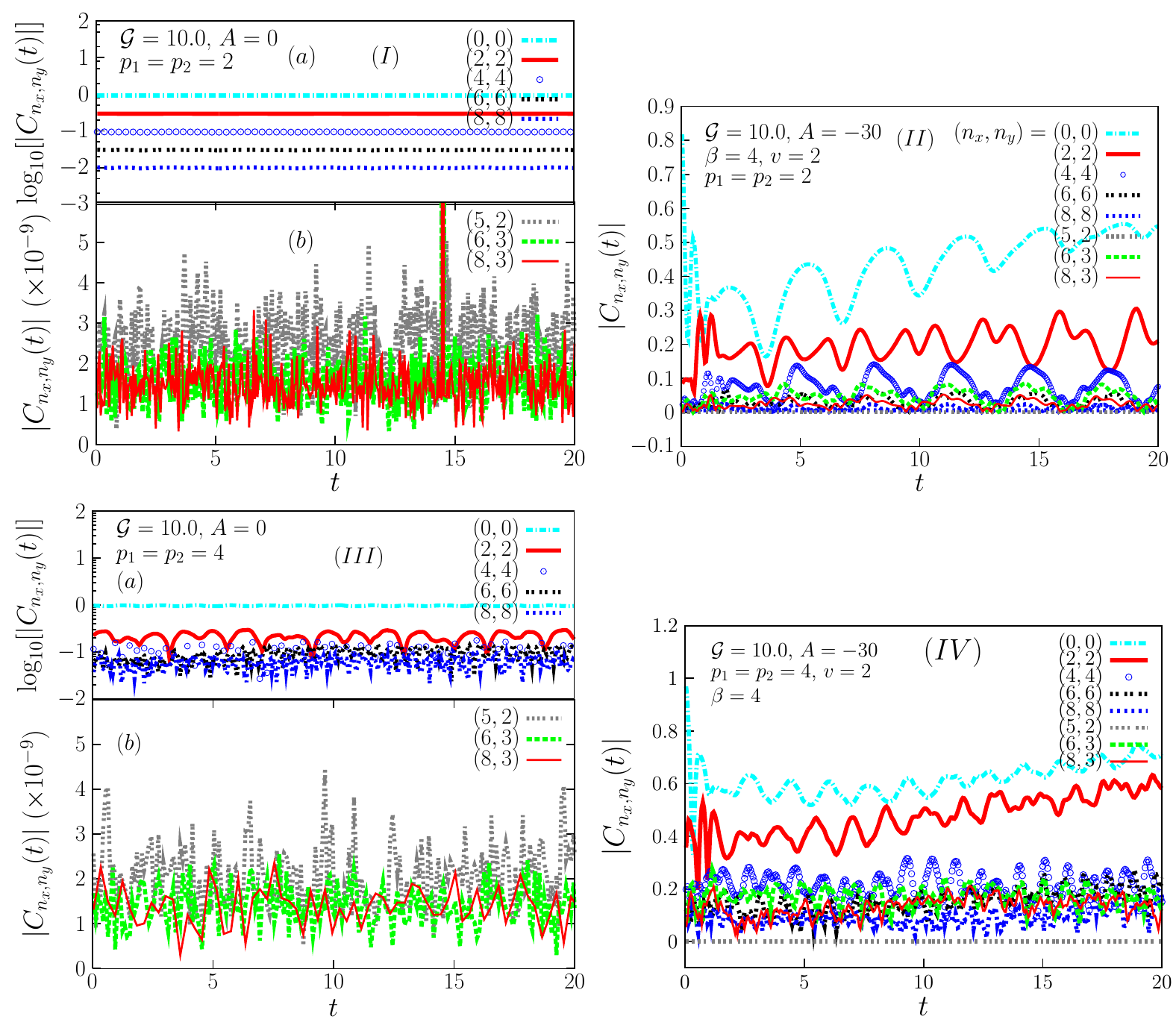}
\caption{(Color online) Dynamics of the weights of the harmonic oscillator states, $|C_{n_x,n_y}(t)|$ of 
Eq.~(\ref{eq:Cnxny}). In frame ($I$), the system is the same as in 
Fig.~\ref{fig:ComponentsplotEvstG10B4T100Step5percentKseveralA}; but for $A=0$, i.e., no applied laser. 
Subframe ($a$) displays $\log_{10}[|C_{n_x,n_y}(t)|]$: dashed-dotted line: $(n_x,n_y)=(0,0)$; solid line: 
$(2,2)$; open circles: $(4,4)$; double-dotted line: $(6,6)$; dotted line: $(8,8)$. Subframe ($b$) displays 
$|C_{n_x,n_y}(t)|$ in units of $10^{-9}$: quadro-triple dotted line: $(5,2)$; dashed line: $(6,3)$; and 
thin solid line: $(8,3)$. Frame ($II$) is the same as ($I$) with the same labels; but with a laser applied 
of parameters $A=-30$, $\beta=4$, and $v=2$. Frame ($III$) is the same as ($I$) with the same labels; but 
for $p_1=p_2=4$. The lower frame ($III,b$) displays $|C_{n_x,n_y}(t)|$ in units of $10^{-9}$. Frame ($IV$) 
is the same as ($III$) with the same labels; but with a laser of depth $A=-30$. $A$ is in units of \hw, 
${\cal G}$ in $(\sqrt{2}a_{ho}^2)^{-1}$, $v$ in $a_{ho}$, $\beta$ in $(a_{ho})^{-2}$, and $t=\tau\omega_{ho}$ 
is unitless.}
\label{fig:plotweightofHOstatesFigureStack}
\end{SCfigure*}%Fig6

\section{Analysis of results}\label{sec:analysis-and-proof-of-chaos}

\subsection{Physical versus numerical chaos}

\hs Br{\'e}zinova {\it et al.} {\cite{Brezinova:2011} have devised powerful tests for detecting the 
presence of numerical chaos that could result from the nonintegrability of the TDGPE. Two of these tests 
are: (1) the conservation of total energy and (2) time-reversal propagation. So far it has already been 
demonstrated \cite{Sakhel:2013} that the total energy of our systems is conserved once the stirrer has 
left the trapping area. For test (2), a time-reversed evolution of the system has been performed that 
begins at some chosen time $t_0$ along the energy dynamics after the stirrer has left the BEC and the 
energy has stabilized. Let us consider a wavefunction $\psi_\pm(\mathbf{r},t)$, which evolves 
``forward" $(+)$/``backward" $(-)$ in time. Once $\psi_+(\mathbf{r},t)$ has evolved from a time $t$ 
to $t_0$, its propagation is reversed from $t_0$ to $t^{\prime}$ (where $t_0>t^{\prime}$) according to
\begin{equation}
\psi_-(\mathbf{r},t^\prime)=U_-(t^\prime,t_0)\psi_+(\mathbf{r},t_0),
\end{equation}

where $U_-$ is the propagator that reverses the evolution of $\psi_+(\mathbf{r},t)$. If the forward 
evolving $\psi_+(\mathbf{r},t)$ equals the backward evolved $\psi_-(\mathbf{r},t^\prime)$ at some 
common time $t=t^\prime$, then numerical chaos is excluded. For a graphical comparison between forward- 
and reversed-evolving properties, the former are plotted from $t=0$ to $t=t_0$, and the latter against 
$t=2 t_0-t^\prime$ from $t=t_0$ to $t=2 t_0$. That is, $t\in [0,t_0]$ corresponds to increasing time 
for $\psi_+(\mathbf{r},t)$ and $t \in [t_0,2 t_0]$ corresponds to decreasing times $t^\prime \in [t_0,0]$ 
for $\psi_{-}(\mathbf{r},t^\prime)$.

\hs Figure~\ref{fig:plottimereversaltestG10A-30V2B4powx2powy2Energy} displays this comparison where it can 
be seen that the dynamics are exactly symmetric about the $t=t_0$ axis. From this, we conclude that any 
chaos demonstrated in this work is physical and not numerical.

\subsection{Chaos and order in the weights of bases states}\label{sec:HO-states}

\hs The goal now is to search for signals of chaos in the GP wavefunction via the mode expansion 
Eq.~(\ref{eq:wave-function-expansion-into-HO-states}). Upon excitation, the trapped Bose gas is energized 
to a number of HO states. The transitions between these states tend to be irregular and it turns out that 
this is one source of chaos in the wavefunction. It must therefore be emphasized that any chaos appearing 
in the total wavefunction $\varphi(x,y;t)$ is translated to chaos in the observables under current study 
[Eqs.~(\ref{eq:Ezp})-(\ref{eq:rms-radius})]. Within this context, Fig.~\ref{fig:plotweightofHOstatesFigureStack} 
displays the dynamics of $|C_{n_x,n_y}(t)|$. Four cases are considered: two for a BEC in a harmonic trap, 
which evolve in the presence and absence of a laser, and likewise two for a BEC in an anharmonic trap. 
From this, the dynamics of $|C_{n_x,n_y}(t)|$ gives a measure for the frequency of particle transitions 
from one HO state to another and indicates whether one has chaos or order. Frame ($I$) refers to a BEC in 
a harmonic trap without a laser. The $|C_{n_x,n_y}(t)|$ with ($n_x,n_y$) even numbers are almost constant 
with time and their values range from order $\sim 10^{-2}$ to order 1. This constancy indicates order 
in $\varphi(x,y;t)$. The largest occupancy is, as expected, for the $(0,0)$ state and some excitations to 
higher states with even $(n_x,n_y)$ are due to the initialization of the system in CN simulations 
\cite{Sakhel:2011,Sakhel:2013}. $|C_{n_x,n_y}(t)|$ in frame ($b$) with either $n_x$ or $n_y$ odd are 
negligible of order $\sim 10^{-9}$ and are rather noisy. These oscillations are however insignificant 
because their weights are vanishingly small. The difference in weights between frames ($a$) and ($b$) is 
attributed to the following: In the absence of a stirrer, $\varphi(x,y;t)$ is even, i.e., symmetric about 
the $x$ and $y$ axes. Therefore, only even HO functions [even $H_{n_x}(x)$ and $H_{n_y}(y)$] contribute to 
the dynamics; the odd functions yield almost zero contribution such that $|C_{n_x,n_y}(t)|\rightarrow0$. 
Frame ($II$) refers to a BEC in a harmonic trap with a laser. In this, $|C_{n_x,n_y}(t)|$ reveals nearly 
periodic oscillations for almost all $(n_x,n_y)$ and range from order $10^{-2}$ up to $10^{-1}$. However, 
this is not enough to cause ordered oscillations in the right column of 
Fig.~\ref{fig:ComponentsplotEvstG10B4T100Step5percentKseveralA}. Because $\varphi(x,y;t)$ is now antisymmetric 
about the $x$-axis, states with odd $n_x$ or $n_y$ display now larger weights than in frame ($I$,$b$) when 
they were practically unoccupied giving only a noisy pattern.

\begin{figure}
\includegraphics[width=8.0cm,viewport=59 130 323 728,clip]{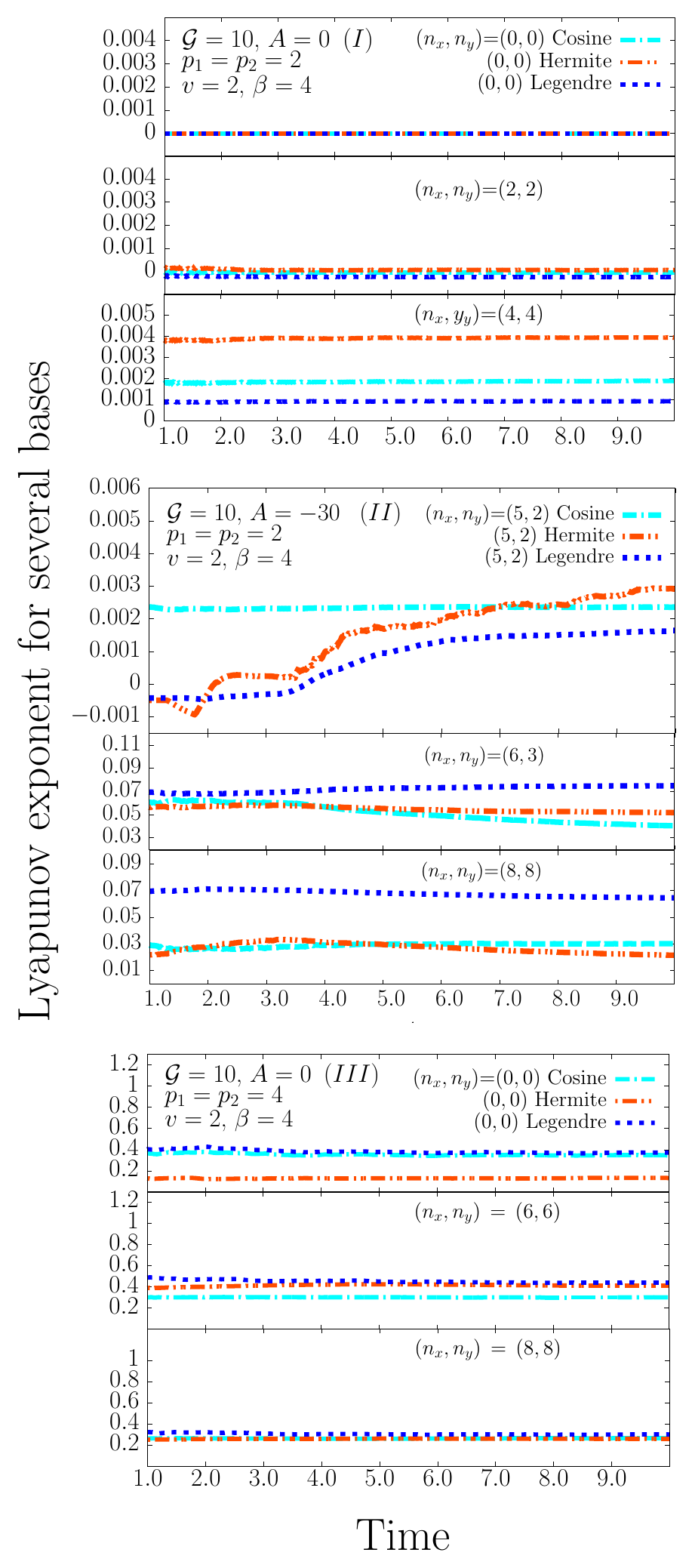}
\caption{(Color online) Lyapunov exponents of the weights $|C_{n_x,n_y}(t)|$ [Eq.~(\ref{eq:Cnxny})] for different bases used in 
the expansion of the total wavefunction. The $x-$axis is the time $t$ in units of $10^4$. The system is in 
principle that of Fig.~\ref{fig:ComponentsplotEvstG10B4T100Step5percentKseveralA}. Frame ($I$) is for $A=0$ 
and $p_1=p_2=2$. Dashed-dotted line: Cosine basis; dashed double-dotted line: Hermite; and dotted line: 
Legendre. Top subframe: state $(n_x,n_y)=(0,0)$; middle: $(2,2)$; and bottom: $(4,4)$. Frame ($II$) is for 
$A=-30$ and $p_1=p_2=2$. Legends are the same as in ($I$). Top frame: $(5,2)$; middle: $(6,3)$; bottom: $(8,8)$. 
Frame ($III$) is for $A=0$ and $p_1=p_2=4$. Legends are again as above. Top subframe: (0,0); middle: (6,6); and 
bottom: (8,8). $A$ is in units of $\hbar\omega_{ho}$, ${\cal G}$ in $(\sqrt{2} a_{ho}^2)^{-1}$; $\beta$ in 
$(a_{ho}^{-2})$; and $t=\tau\omega_{ho}$ is unitless.}
\label{fig:plotlyapmaxSeveralCasesCrankNicolson}
\end{figure}%Fig7

\hs Frame ($III$) refers to a BEC in an anharmonic trap without a laser. In ($a$), $|C_{0,0}(t)|$ is almost 
constant in the absence of a laser. $|C_{2,2}(t)|$ oscillates almost regularly whereas the $|C_{n_x,n_y}(t)|$ 
for (4,4), (6,6), and (8,8) are now chaotic. Indeed, even if only one state $(n_x,n_y)$ is chaotic, all the 
dynamical quantities that can be obtained from $\varphi(x,y;t)$ will be chaotic. In frame ($b$), (5,2), 
(6,3), and (8,3) show a noisy pattern as in frame ($I$,$b$) of negligible amplitude but are not expected 
to contribute to the overall dynamic behavior of $\varphi(x,y;t)$. Frame ($IV$) is for a BEC in an anharmonic 
trap with a laser. In this, some ordered oscillations appear in $|C_{0,0}(t)|$, $|C_{2,2}(t)|$, and $|C_{4,4}(t)|$ 
whereas $|C_{5,2}(t)|$ seems to be almost zero. The remainder of $|C_{n_x,n_y}(t)|$ are chaotic and their 
values are largely of order $\sim 10^{-2}$. The latter can be related to the chaos displayed in the right 
column of Fig.~\ref{fig:ComponentsplotEvstG10B4T30Step5percentK1severalp}. Therefore, the anharmonic trap 
yields irregular oscillations in $|C_{n_x,n_y}(t)|$ that manifest as chaotic oscillations in the physical 
observables.

\subsection{Lyapunov exponents in different bases}

\hs The unique signature for chaos in the expansion coefficients $C_{n_x,n_y}(t)$ is again a positive ${\cal L}$. 
Hence, we analyze chaos in the GP wavefunction itself by expanding it in different bases to examine whether chaos 
in $|C_{n_x,n_y}(t)|$ is basis invariant. Fig.~\ref{fig:plotlyapmaxSeveralCasesCrankNicolson} displays ${\cal L}$ 
for the $|C_{n_x,n_y}(t)|$ evaluated via a Cosine, Hermite, and Legendre basis. In Frame ($I$), there is practically 
no difference in the qualitative behavior of ${\cal L}$ for the state $(n_x,n_y)=(0,0)$ between the different bases. 
${\cal L}$ goes to zero in all bases and therefore chaos is absent in all of them. For $(2,2)$, ${\cal L}$ is positive and 
close to zero. It varies slightly with bases; but it still gives the same qualitative result indicating order. For 
the state $(4,4)$, the values of ${\cal L}$ vary with bases; but they are all positive signaling chaos.

\hs In Frame ($II$), ${\cal L}$ for (5,2) in the Cosine basis is positive and almost constant in the time-range 
considered. However, for the Hermite and Legendre bases, ${\cal L}$ evolves from a negative towards a positive 
value. That is, for (5,2) $|C_{n_x,n_y}(t)|$ is eventually chaotic for all bases as $t\rightarrow\infty$. For (6,3) 
and (8,8), ${\cal L}$ displays positive values for all bases and $|C_{n_x,n_y}(t)|$ is chaotic. In Frame ($III$), 
${\cal L}$ is positive in all bases for (0,0), (6,6), and (8,8). Therefore, the result of having chaos in the GP 
wavefunction that is basis independent can be considered as a new test for the presence of physical chaos taking 
into account that the three bases we are using are vastly different functions!

\subsection{Chaos in coordinate space}

\hs Figure~\ref{fig:plotLyapunovAnharmonicHarmonicArtificialHbar} further confirms the presence of spatial chaos 
in the coordinate space of the BEC via the evolution of ${\cal L}$ for $R_{rms}$. The computations ran for a time 
$t=10000$, long enough to examine the asymptotic-time behavior of ${\cal L}$. For the upper and lower frames, the 
parameters of the time series analysis and the resulting asymptotic Lyapunov exponents ${\cal L}_{asymp}$ are listed 
in Tables~\ref{tab:tabLyapunov1} and \ref{tab:tabLyapunov2}, respectively. All ${\cal L}$ converge to stable positive 
values after a long simulation confirming the existence of chaos. The stability of ${\cal L}$ signals that once 
chaos has been initiated in a BEC, it does not decay if one allows the BEC to evolve for a long time. Therefore, 
one needs to design ways to suppress chaos, particularly if it arises in quantum computations.

\begin{table}[t!]
\caption{Asymptotic Lyapunov exponents ${\cal L}_{asymp}$ of $R_{rms}$ for the systems of 
Figs.~\ref{fig:ComponentsplotEvstG10B4T100Step5percentKseveralA} and \ref{fig:ComponentsplotEvstG10B4T30Step5percentK1severalp}. 
Parameters are: trapping exponents $p_1$ and $p_2$ and stirrer depth or height $A$ [see Eq.~(\ref{eq:combined-potential})], 
optimal embedding delay $\tau$, and minimal required dimension $m$. The velocity of the stirrer is $v=2$ and the parameter 
describing its width is $\beta=4$. $A$ is in units of \hw, $\beta$ in $a_{ho}^{-2}$, and $v$ in $a_{ho}$.}
\begin{center}
\begin{tabular}{*{6}{@{\hspace{0.3cm}}c@{\hspace{0.3cm}}}} \hline
$p_1$ & $p_2$ & $A$ & $\tau$ & $m$ & ${\cal L}_{asymp}$ \\ \hline\hline
2 & 2 & $+20$ & 16 & 5 & 0.056033 \\
2 & 2 & $+30$ & 16 & 5 & 0.081165 \\
2 & 2 & $+40$ & 15 & 5 & 0.081559 \\
2 & 2 & $-20$ & 15 & 5 & 0.280095 \\
2 & 2 & $-30$ & 15 & 5 & 0.076887 \\
2 & 2 & $-40$ & 14 & 5 & 0.015436 \\
2.8 & 2.8 & $-30$ & 11 & 7 & 0.188084 \\
2.8 & 2.8 & $+30$ & 11 & 7 & 0.268660 \\
5 & 5 & $+30$ & 8 & 7 & 0.099709 \\
5 & 5 & $-30$ & 6 & 7 & 0.619132 \\
7 & 7 & $-30$ & 5 & 7 & 0.365345 \\
7 & 7 & $+30$ & 5 & 7 & 0.339876 \\ \hline
\end{tabular}
\end{center}
\label{tab:tabLyapunov1}
\end{table}

\begin{table}[t!]
\caption{As in Table~\ref{tab:tabLyapunov1}; but for the systems of 
Fig.~\ref{fig:QuantumeffectsRMSradiusvariousRelativehbarPropertiesBlueLaser} at various values of the relative 
constant $\Gamma$. The amplitude of the stirrer is fixed at $A=+30$ and $p_1=p_2=2$. Parameters are: $\Gamma$, 
optimal embedding delay $\tau$, and minimal required dimension $m$. $A$ is in units of \hw.}
\begin{center}
\begin{tabular}{*{4}{@{\hspace{0.3cm}}c@{\hspace{0.3cm}}}} \hline
$\Gamma$ & $\tau$ & $m$ & ${\cal L}_{asymp}$ \\ \hline\hline
 4 & 4 & 6 & 0.833164 \\ 
 6 & 3 & 6 & 0.676284 \\ 
10 & 1 & 6 & 0.364494 \\ 
16 & 1 & 6 & 0.571895 \\ 
28 & 1 & 6 & 0.629997 \\ 
40 & 1 & 6 & 1.186200 \\ 
64 & 1 & 6 & 0.272363 \\ \hline
\end{tabular}
\end{center}
\label{tab:tabLyapunov2}
\end{table}

\begin{figure}[t!]
\includegraphics[width=8.5cm,viewport=82 475 361 770,clip]{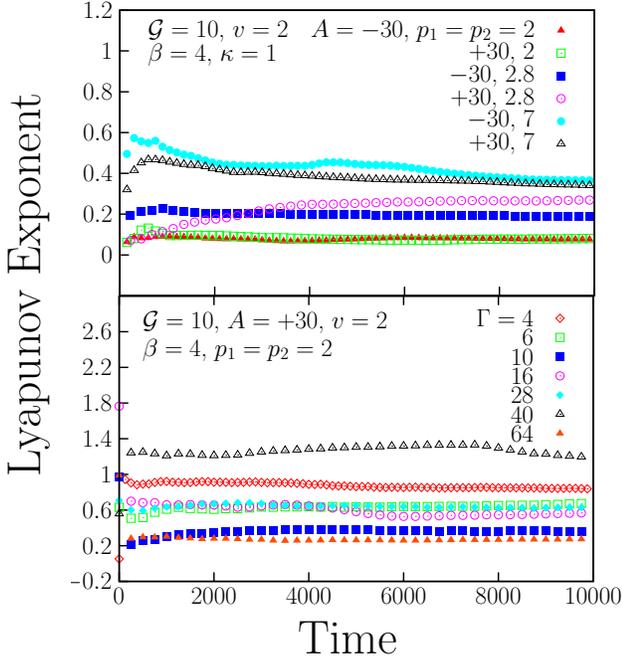}
\caption{(Color online) Evolution of the Lyapunov exponent ${\cal L}$ for various systems considered from 
Figs.~\ref{fig:ComponentsplotEvstG10B4T100Step5percentKseveralA} and \ref{fig:ComponentsplotEvstG10B4T30Step5percentK1severalp} 
after a very long simulation time $t=10000$. The interaction parameter is ${\cal G}=10$ and the velocity of the stirrer 
is $v=2$ with width parameter $\beta=4$. Upper frame: ${\cal L}$ of $R_{rms}(t)$ for various values of $A$ and 
$p_1=p_2$. Solid triangles: $A=-30$, $p_1=p_2=2$; open squares: $+30$, 2; solid squares: $-30$, 2.8; open circles: 
$+30$, 2.8; solid circles: $-30$, 7; open triangles: $+30$, 7. Lower frame: ${\cal L}$ at various values of 
$\Gamma$ [cf. Eq.(\ref{eq:gamma-explanation})] for $A=+30$ and $p_1=p_2=2$. Open diamonds: $\Gamma=4$; open 
squares: $6$; solid squares: $10$; open circles: $16$; solid circles: $28$; open triangles: $40$; and solid 
triangles: $64$. $A$ is in units of \hw, ${\cal G}$ in $(\sqrt{2}a_{ho}^2)^{-1}$, $v$ in $a_{ho}$, $\beta$ 
in $(a_{ho})^{-2}$, and $t=\tau\omega_{ho}$ is unitless.}
\label{fig:plotLyapunovAnharmonicHarmonicArtificialHbar}
\end{figure}%Fig8

\subsection{Chaos in energy space}\label{sec:energy_trajectories}

\hs We now confirm the existence of chaos in energy space for most of the cases considered in this work. 
Fig.~\ref{fig:EnergyChaosG10A30B4V2T30K1powx2powy2} shows trajectories $(E,\dot{E})$ for all energy components 
$E$ in addition to ($R_{rms},\dot{R}_{rms})$ in a harmonic trap stirred by a blue-detuned laser. In this, 
order is categorically demonstrated in $E_{kin}$, $E_{osc}$, and $R_{rms}$ because their trajectories display 
periodic behavior whereas chaos is prevalent in the other components as they are aperiodic. A trend is also 
revealed by all chaotic components in developing an attractor as, for example, the R{\"o}ssler attractor 
\cite{Rossler:1976,Amaral:2006} by $E_{flow}$ and $E_{int}$. The separated-out trajectories demonstrate the 
state of the system while the BDLP is inside the trap.

\begin{figure}[t!]
\includegraphics[width=8.7cm,viewport=76 206 523 716,clip]{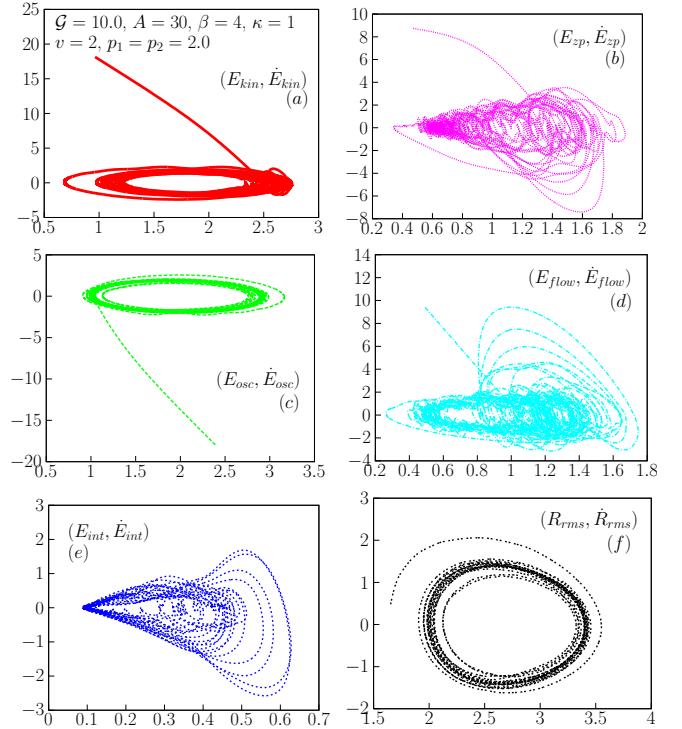}
\caption{(Color online) Energy trajectories $(E,\dot{E})$ of the system in 
Fig.~\ref{fig:ComponentsplotEvstG10B4T100Step5percentKseveralA}[($c$) and ($d$)] for a time of $t=20$. 
Frame ($a$) $E=E_{kin}$; ($b$) $E_{zp}$; ($c$) $E_{osc}$; ($d$) $E_{flow}$; ($e$) $E_{int}$; and ($f$) 
additionally the phase-space trajectory $(R_{rms},\dot{R}_{rms})$. Lengths and energies are in units 
of the trap, $a_{ho}$ and $\hbar\omega_{ho}$, respectively, and $t$ is unitless.}
\label{fig:EnergyChaosG10A30B4V2T30K1powx2powy2}
\end{figure}%Fig9

\begin{figure}[t!]
\includegraphics[width=9.0cm,viewport=71 239 547 763,clip]{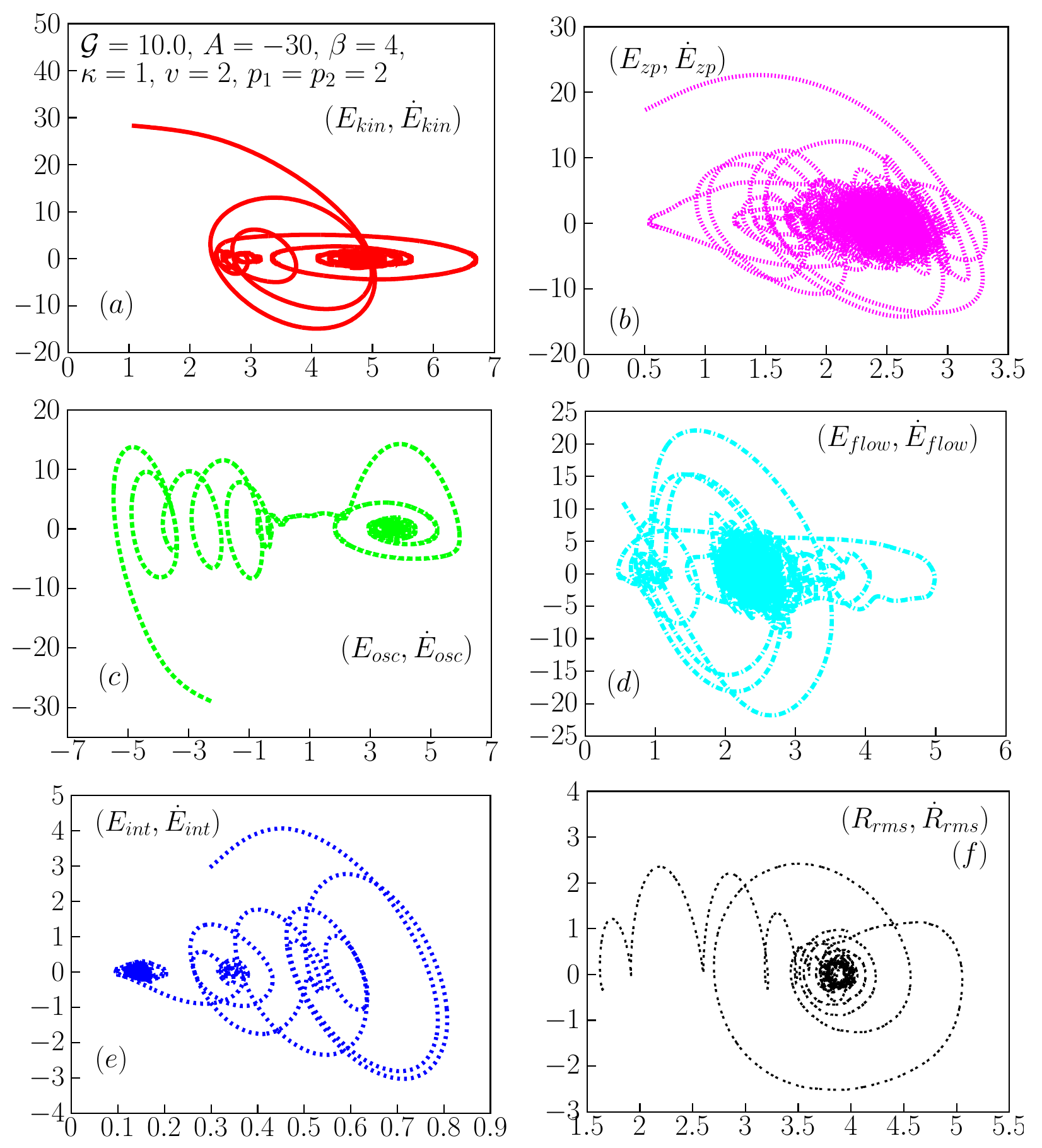}
\caption{(Color online) As in Fig.~\ref{fig:EnergyChaosG10A30B4V2T30K1powx2powy2}; but for $A=-30$ 
(system in Fig.~\ref{fig:ComponentsplotEvstG10B4T100Step5percentKseveralA}[($i$) and ($j$)]).}
\label{fig:EnergyChaosG10A-30B4V2T30K1powx2powy2}
\end{figure}%Fig10

\begin{figure}[t!]
\includegraphics[width=9.0cm,viewport=77 264 528 766,clip]{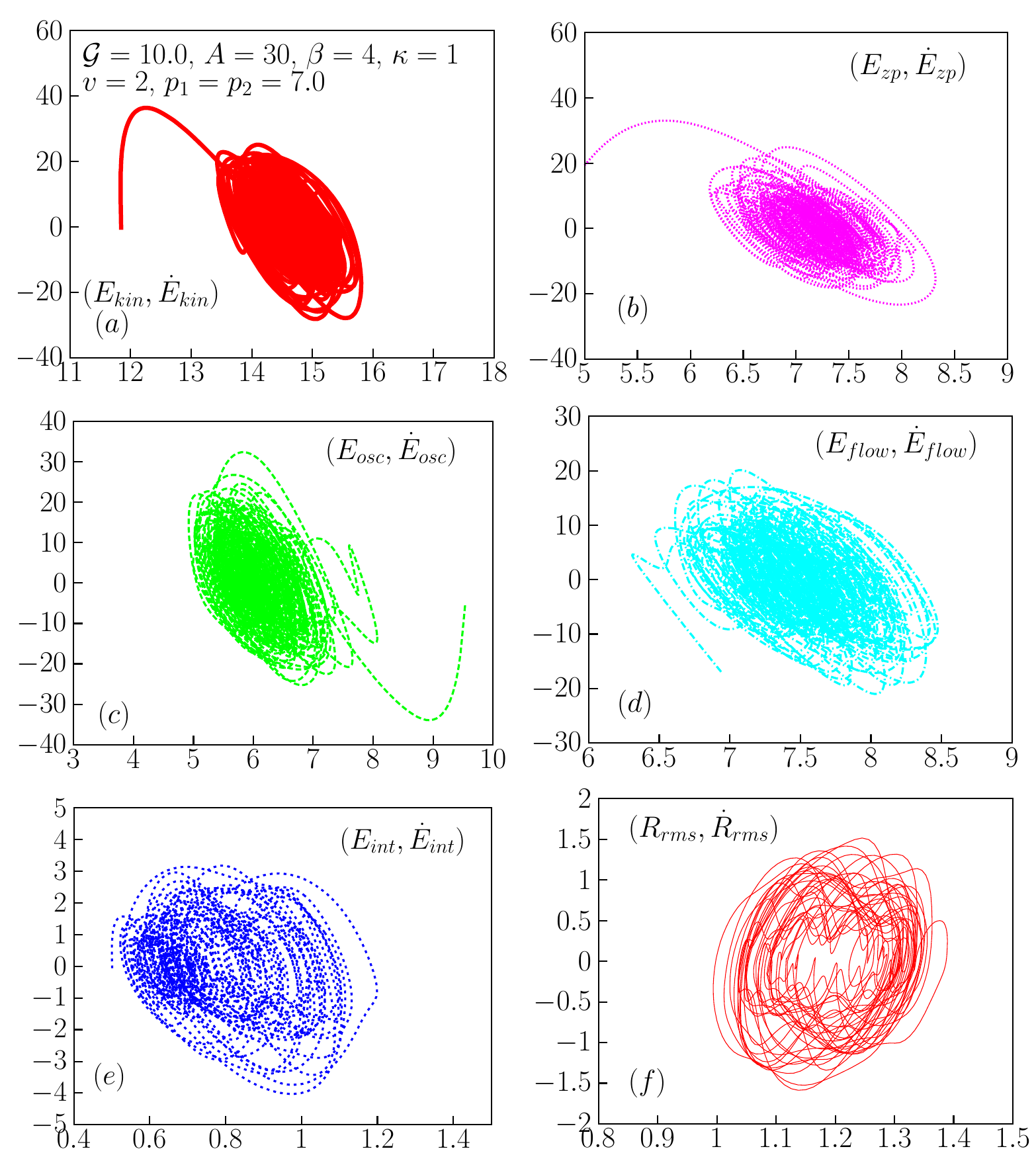}
\caption{(Color online) As in Fig.~\ref{fig:EnergyChaosG10A30B4V2T30K1powx2powy2}; but for $p_1=p_2=7$ 
(system in Fig.~\ref{fig:ComponentsplotEvstG10B4T30Step5percentK1severalp} [frames ($e$) and ($f$)]).}
\label{fig:EnergyChaosG10A30B4V2T30K1powx7powy7}
\end{figure}%Fig11

\begin{figure}
\includegraphics[width=9.3cm,viewport=72 235 546 763,clip]{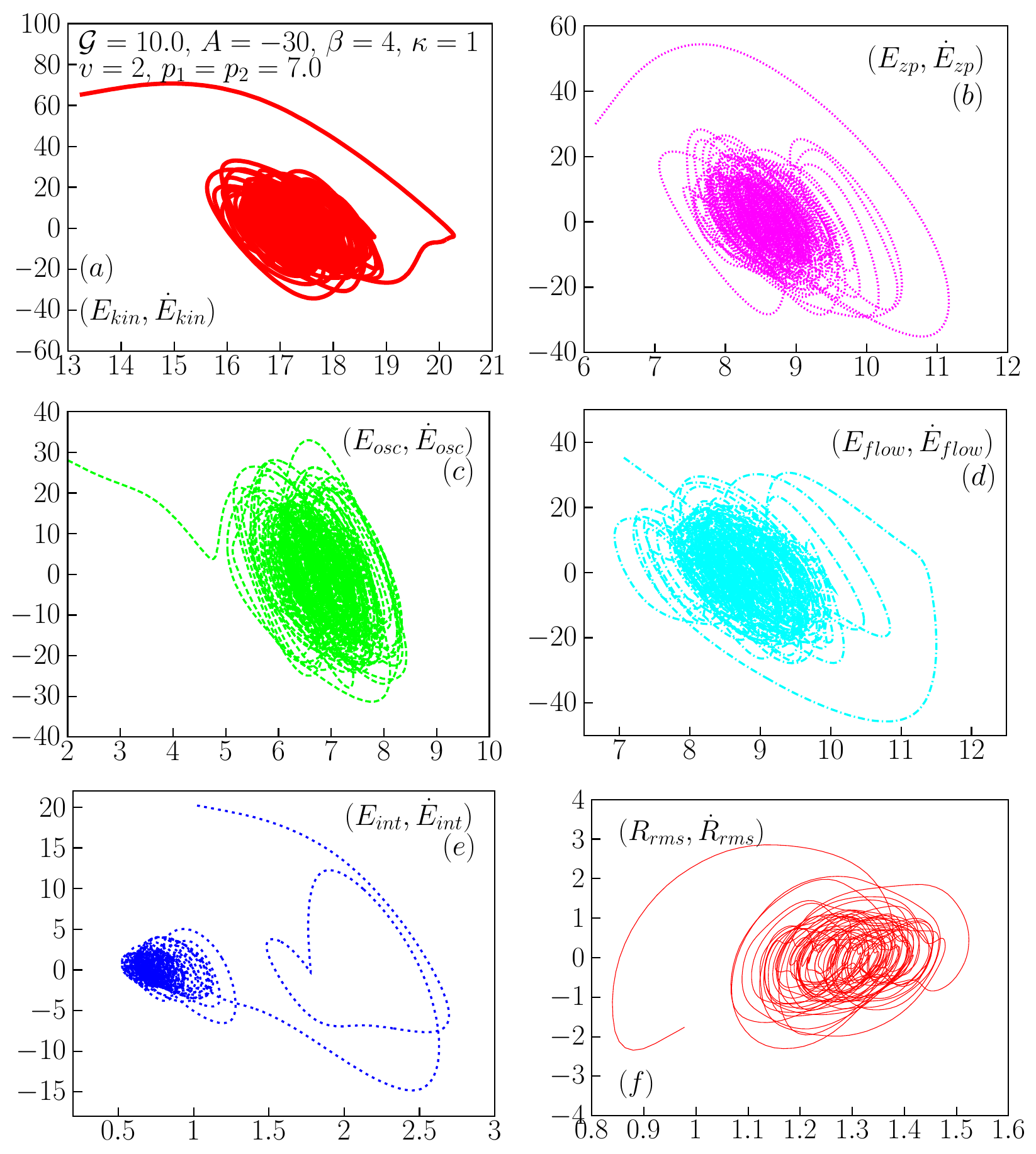}
\caption{(Color online) As in Fig.~\ref{fig:EnergyChaosG10A-30B4V2T30K1powx2powy2}; but for $p_1=p_2=7$ 
(system in Fig.~\ref{fig:ComponentsplotEvstG10B4T30Step5percentK1severalp} [frames ($k$) and ($\ell$)]).}
\label{fig:EnergyChaosG10A-30B4V2T30K1powx7powy7}
\end{figure}%Fig12

\hs Figure\ft\ref{fig:EnergyChaosG10A-30B4V2T30K1powx2powy2} in turn demonstrates that in the same harmonic 
trap but with red-detuned laser stirring, $E_{kin}$ and $E_{osc}$ are no longer ordered and the rest of the 
energy components remain chaotic. It is possible that $R_{rms}$ may be spiraling into a kind of periodic 
behavior at longer times after exhibiting chaos. Indeed, all trajectories seem to evolve towards a concentrated dense 
``area" because once the RDLP leaves the BP, the number of excited states goes down as the points on the trajectories 
come closer together. Indeed, as can be deduced from frame ($f$), the volume of phase-space is reduced as the BEC 
evolves with time indicating a decline in the number of energy states to which the system is excited. Comparing 
Figs.~\ref{fig:EnergyChaosG10A30B4V2T30K1powx2powy2} and \ref{fig:EnergyChaosG10A-30B4V2T30K1powx2powy2}, one can 
see then that the RDLP generates chaos in $E_{kin}$, $E_{osc}$, and $R_{rms}$ because it introduces a larger phase 
space density. Consequently, the extent of the trajectories along both axes is larger for the red than blue-detuned 
laser.

\hs In Fig.~\ref{fig:EnergyChaosG10A30B4V2T30K1powx7powy7}, chaos is also signaled by all the above physical 
observables in an anharmonic trap with a blue-detuned laser. In this case, the trajectories probe more of the 
space of ($E$,$\dot{E}$) and a larger number of excited states is manifested than for the corresponding harmonic 
trap in Fig.~\ref{fig:EnergyChaosG10A30B4V2T30K1powx2powy2}. This is because anharmonicity introduces a larger 
quantum pressure due to stronger confinement that excites the BEC to higher energy levels. In the corresponding 
Fig.~\ref{fig:EnergyChaosG10A-30B4V2T30K1powx7powy7} with a red-detuned laser, the patterns of the trajectories 
are qualitatively not very different from their counterparts in Fig.~\ref{fig:EnergyChaosG10A30B4V2T30K1powx7powy7} 
demonstrating that the trap anharmonicity has become dominant in determining the qualitative behavior of the trajectories. 
The trajectories $(R_{rms},\dot{R}_{rms})$ in Figs.~\ref{fig:EnergyChaosG10A30B4V2T30K1powx7powy7} and 
\ref{fig:EnergyChaosG10A-30B4V2T30K1powx7powy7} clearly exhibit chaos as they do not form periodic orbits. Compared 
with the harmonic trap, the extent of these trajectories along $R_{rms}$ is reduced, whereas along $E$ it increases 
with increasing confinement strength via $p_1=p_2$.

\subsection{More quantum effects}

\hs What remains now is to examine chaos with increased $\Gamma$. Fig.~\ref{fig:EnergyChaosG10A30B4V2T30K1powx2powy2SeveralBeta} 
shows the trajectories ($X$, $\dot{X}$) for the evolution of $R_{rms}$, $E_{zp}$, and $E_{flow}$. These figures verify the 
presence of chaos in coordinate and energy space with increased $\Gamma$ because none of them manifests periodic behavior. 
The range of these observables and their time-derivatives rises with increasing $\Gamma$. For $R_{rms}$, this shows that an 
increased degree of chaos implies a rise in the relative sizes of the BEC and external trap. One can also ascribe to this 
behavior an artificial rise in the ``volume" of phase-space that signals an increase in the number of energy states. This 
is further supported by an increase in the range of $E_{zp}$ and $E_{flow}$ (and their derivatives).

\begin{SCfigure*}
\includegraphics[width=11.0cm,viewport=73 257 570 765,clip]{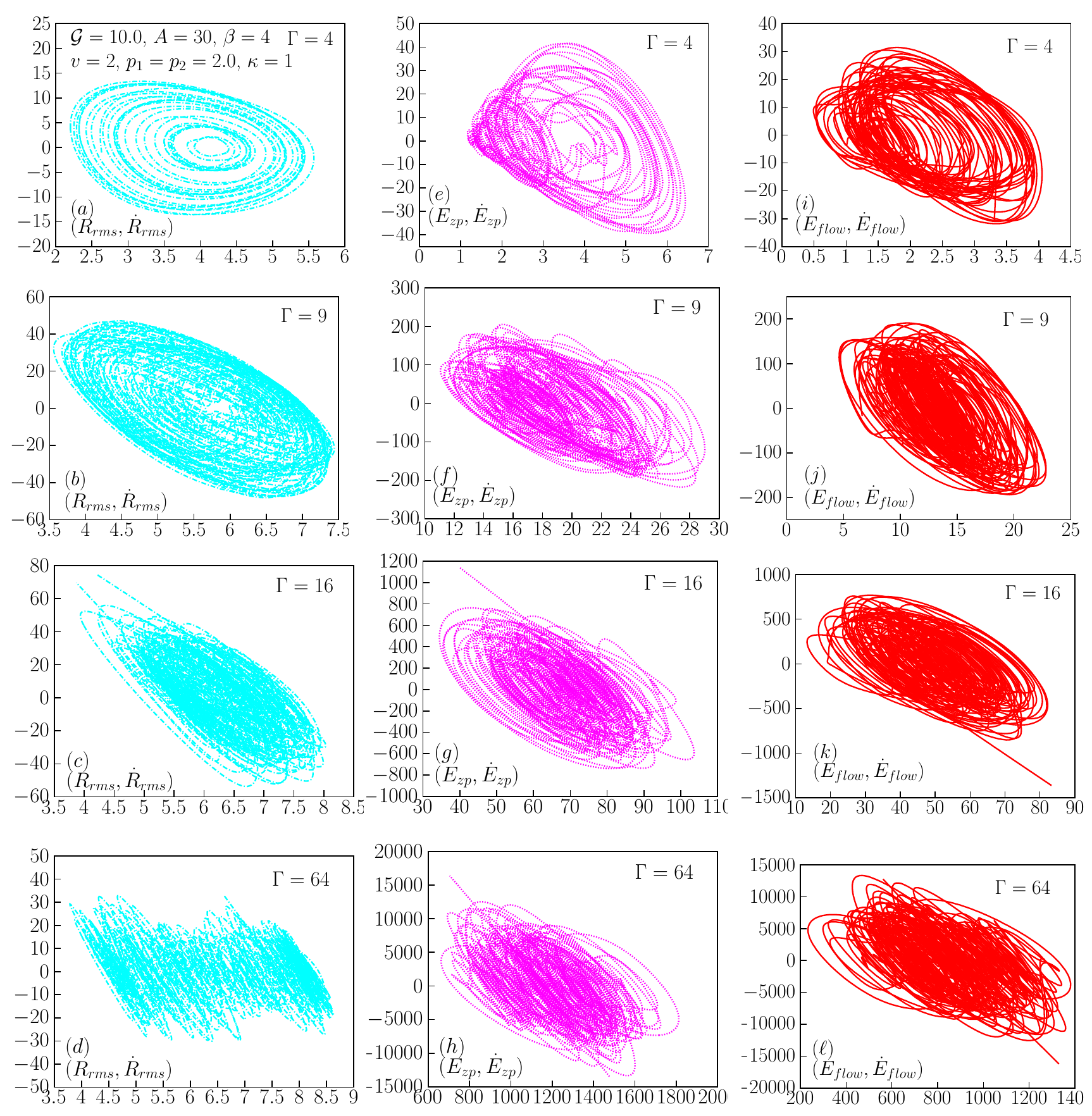}
\caption{(Color online) Quantum effects revealed by the phase-space and energy trajectories for the system of 
Fig.~\ref{fig:QuantumeffectsRMSradiusvariousRelativehbarPropertiesBlueLaser}. In all frames, the relative 
Planck's constant $\Gamma$ is increased from top to bottom according to the sequence 4, 9, 16, and 64. Left 
column [frames ($a-d$)]: phase-space trajectories $(R_{rms},\dot{R}_{rms})$; middle column [frames ($e-h$)]: 
zero-point energy trajectories $(E_{zp},\dot{E}_{zp})$; right column [frames ($i-l$)]: kinetic-flow energy 
trajectories $(E_{flow},\dot{E}_{flow})$. $R_{rms}(t)$ is in units of $a_{ho}$ and $E_{zp}(t)$ and $E_{flow}$ 
are in units of \Rhw.}
\label{fig:EnergyChaosG10A30B4V2T30K1powx2powy2SeveralBeta}
\end{SCfigure*}%Fig13

\subsection{Confirmation of energy chaos via the Lyapunov exponent}

\hs The aim now is to apply the Lyapunov exponent as a measure that further confirms the presence or absence 
of chaos and how long it persists in the BEC. In Fig.~\ref{fig:EnergyChaosLyapunovExponentsSeveralSystems}, 
the evolution of ${\cal L}$ for the energy components is displayed. Scanning all frames, it can be seen that 
after a long simulation time ${\cal L}$ is positive for all the observables under consideration; except in 
frame $(K)$. However, it was found hard to assign a certain behavioral pattern for ${\cal L}$ in terms of 
trapping geometry and laser parameters. For a harmonically trapped BEC that is excited by a red-detuned laser, 
${\cal L}$ for the energies seems to approach stable values, except for $E_{zp}$ where ${\cal L}$ declines after 
$t\sim 2000$. This decline is an indication that the degree of chaos in $E_{zp}$ decreases with time. Similarly, 
with a blue-detuned laser, the ${\cal L}$'s stabilize with time except for $E_{flow}$ and $E_{int}$, which keep 
rising, and with it the degree of chaos in them. For anharmonic trapping with a red- or blue-detuned laser, 
there is no qualitative change in the behavior of ${\cal L}$ when compared with the corresponding harmonic 
trapping. For some of the observables, the values of ${\cal L}$ are significantly larger than for the harmonic 
trapping. For example, in frame ($F$), ${\cal L}$ for $E_{kin}$ reaches $\sim 3.6$ whereas in frame ($B$) it is 
$\sim 0.2$ for the same blue-detuned laser. In frame ($E$), ${\cal L}$ for $E_{flow}$ reaches $\sim 4.6$ compared 
with $\sim 3.5$ in frame ($A$) for the same red-detuned laser, and similarly for other observables. Over some 
energy intervals, chaos arises with increasing trapping anharmonicity bringing this in line with the behavior 
of energy trajectories in Figs.~\ref{fig:EnergyChaosG10A30B4V2T30K1powx2powy2}--\ref{fig:EnergyChaosG10A-30B4V2T30K1powx7powy7}. 
Frames ($G$) and ($H$) present the same qualitative information as in frame ($B$) with different blue-detuned 
laser amplitudes. Nevertheless, some predictability can be assigned to the response of the magnitude of ${\cal L}$ 
to increasing quantum effects. ${\cal L}$ for $E_{flow}$ and $E_{zp}$ is seen to rise with increasing $\Gamma$ 
beyond 1. In contrast, with $\Gamma\ll 1$, ${\cal L}$ tends to approach zero except for $E_{osc}$ where ${\cal L}$ 
becomes negative signalling the absence of chaos. This is brought in line with 
observations in Fig.~\ref{fig:QuantumeffectsRelativeHbarG10A30B4T20Beta0.1stack}$(a)$ and $(b)$. Therefore, the 
reduction of quantum effects leads to ordered behavior in physical observables. The chaotic behavior remains in 
general unpredictable. From the previous displays, one concludes that chaos persists for a very long time and does 
not easily vanish in nondissipative systems.

\begin{SCfigure*}
\includegraphics[width=12cm,viewport= 72 172 579 761,clip]{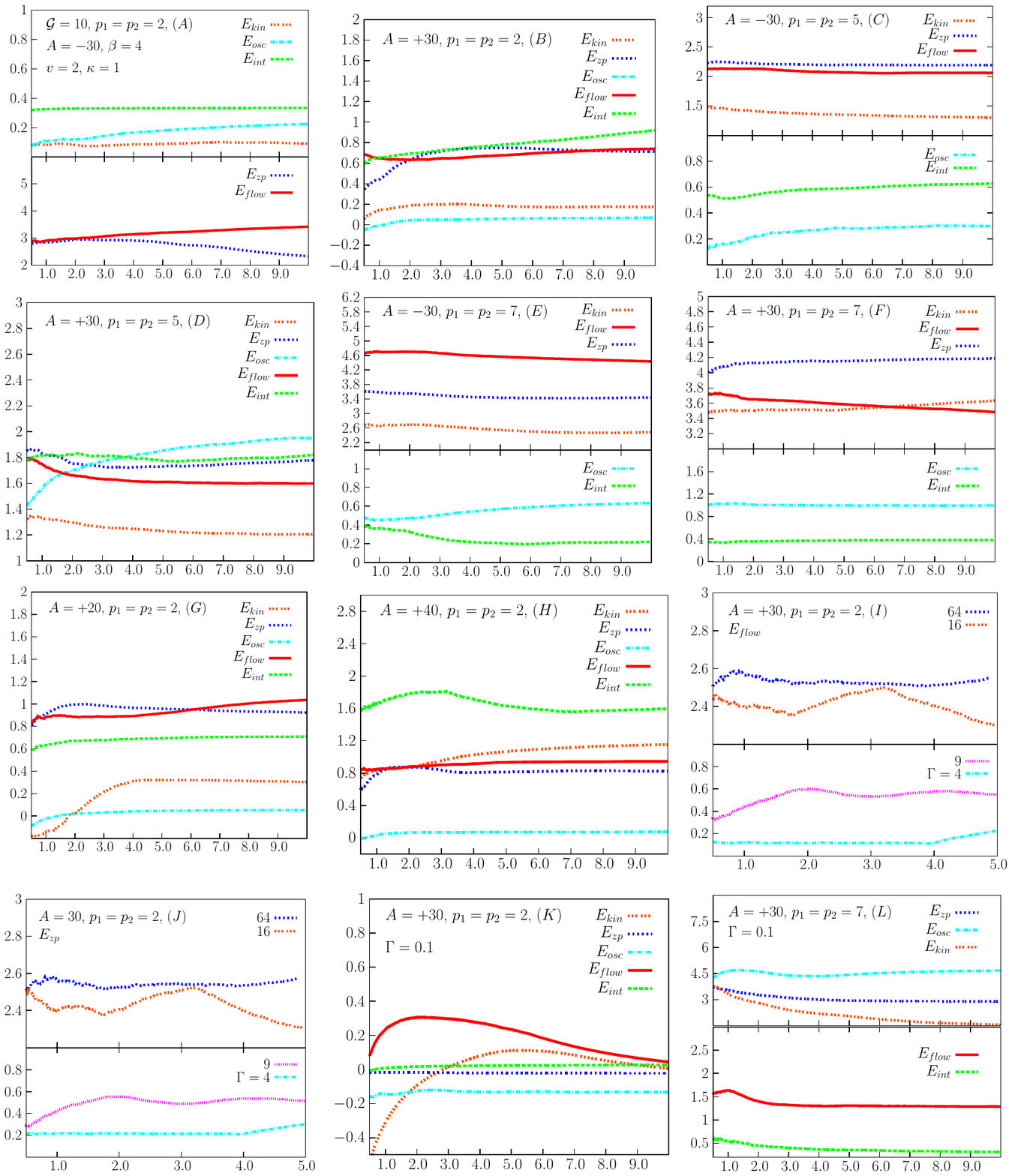}
\caption{(Color online) Lyapunov exponents for the energy dynamics of all the previous systems in 
Figs.~\ref{fig:ComponentsplotEvstG10B4T100Step5percentKseveralA}--\ref{fig:QuantumeffectsRMSradiusvariousRelativehbarPropertiesBlueLaser} 
as a function of evolution time. Energy is along the $y$-axis and time along the $x$-axis. ${\cal G}=10$, $v=2$, 
$\beta=4$ and $\kappa=1$. Frame $(A)$: $A=-30$, $p_1=p_2=2$; $(B)$ $+30$, $2$; $(C)$ $-30$, $5$; $(D)$ $+30$, $5$; 
$(E)$ $-30$, $7$; $(F)$ $+30$, $7$; $(G)$ 20, 2; $(H)$ $+40$, 2. Frames $(I)$ and $(J)$ display the effects of 
increasing the relative Planck's constant $\Gamma$ for $A=+30$ and $p_1=p_2=2$. Dotted line: $\Gamma=64$; 
triple-dotted line: 16; fine-dotted line: 9; and dashed-dotted line: 4. Frames $(K)$ and $(L)$ display the 
reduced quantum effects with $\Gamma=0.1$. Frame $(K)$ $A=+30$, $p_1=p_2=2$; $(L)$ +30, 7. Frames ($A$-$H$), 
$(K)$, and $(L)$ all have the same labels although the locations of the observables in the frames tend to vary: 
triple-dotted line: $E_{kin}$; dashed-dotted line: $E_{osc}$; dashed line: $E_{int}$; dotted line: $E_{zp}$; 
solid line: $E_{flow}$. The time is in units of $10^4$.}
\label{fig:EnergyChaosLyapunovExponentsSeveralSystems}
\end{SCfigure*}%Fig14

\subsection{Analogy to the chaotic billiard}\label{sec:chaotic-billiard}

\hs The spatial chaos and order found in Figs.~\ref{fig:ComponentsplotEvstG10B4T100Step5percentKseveralA}
and \ref{fig:ComponentsplotEvstG10B4T30Step5percentK1severalp} can be explained by the chaotic billiard 
concept \cite{Wimberger:2014} whose effects can be mimicked by an anharmonic trap. In 2D, the trajectories 
of a particle moving with constant energy on a billiard table with defocusing boundaries are chaotic unlike 
one that has a circular-shaped boundary \cite{Wimberger:2014}. For a harmonic trap in 2D, the BEC with low 
kinetic energy is unable to surmount the barrier of the external trap and remains therefore far away from 
the BP hard walls. Accordingly, it oscillates periodically inside a circular area and endures no chaotic 
billiard effect. If the energy is increased, the BEC oscillates within a larger circular area which if 
cutoff by the BP becomes, for certain energy levels, square-like with rounded corners. By increasing 
the energy, the BEC moves up the potential barrier of the external trap and eventually becomes a squared 
area. After this, it no longer oscillates periodically. For an anharmonic trap, the shape of the spatial 
boundary is not circular as for a harmonic trap, but square-like with rounded corners and becomes square 
with increasing anharmonicity. Therefore, the dynamics of the BEC becomes chaotic with a broad excitation 
spectrum and spatial chaos increases with growing anharmonicity. One understands now why there is spatial
order for a blue-detuned laser: For the BDLP parameters considered, the BEC is not excited to energy levels 
where the BP begins to be assertive; the BEC remains therefore inside a circular boundary and oscillates 
periodically. In contrast, the RDLP excites the BEC to high energy levels where the BP begins to be assertive.

\section{Validity of the GPE in the present approach}\label{sec:validityGPE}

\hs Our use of the GPE is justified (1) by the success of previous similar work 
\cite{Horng:2009,Caradoc:2000,Sasaki:2010,Fujimoto:2010,Fujimoto:2011}, (2) because the temperature 
remains in a regime well below the critical temperature $T_c$, and (3) because the scattering length 
$a_s \ll \overline{d}$, where $\overline{d}=\sqrt{L^2/N}=2.818\,\mu m$ is less than the average 
interparticle separation; i.e., our systems are very dilute Bose gases. Regarding (1), Fujimoto and 
Tsubota (FT) \cite{Fujimoto:2011} examined the dynamics of a trapped BEC induced by an oscillating 
Gaussian potential. Their study was based on a numerical simulation of the 2D GPE. Because it was 
thought that the oscillating potential might induce some heating effects that might invalidate their 
use of the GPE, they calculated the increase of temperature and showed that it remained relatively 
very small.

\hs Further support for our arguments can be drawn from the work of Ref.~\cite{Schreck:2001}, where a 
quasipure condensate was identified because it constituted a fraction of only 77$\%$. In addition, the 
GPE was applied to explore BECs excited by obstacles. For example, Sasaki {\it et al.} \cite{Sasaki:2010} 
explored vortex shedding from an obstacle moving inside the BEC. Horng {\it et al.} \cite{Horng:2009} 
examined the dynamics of turbulent flow in a 2D trapped BEC. Caradoc-Davis {\it et al.} \cite{Caradoc:2000} 
simulated the effects of rotationally stirring a 3D trapped BEC with a Gaussian laser beam.

\hs Concerning point (2), FT used the specific-heat equation of the 2D ideal trapped Bose gas written

\begin{equation}
C(T)\,=\,\frac{6 k_B^3 T^3 \xi(3)}{\hbar^2 \omega_x \omega_y},
\label{eq:specific-heat-ideal-Bose-gas}
\end{equation}

to estimate the heating of their condensate. [Here $\xi(n)$ is the Riemann Zeta function with $\xi(3)=1.2021$]. 
This was simply obtained from a division of the energy rise $\Delta E$ by the value of $C(T)$ at the transition 
temperature $T_c$. Because our systems are dilute ($N\sim 117$, $a_s\ll \overline{d}$) and weakly-interacting, 
we can follow FT and apply Eq.~(\ref{eq:specific-heat-ideal-Bose-gas}) to estimate the temperature of our systems 
in the excited state only after the stirrer has left the BP. First of all, for our harmonically trapped BEC 
without excitation by any laser, the energy per particle is $E(A=0)\,=\,1.3340$ (\hw); for the same system, but 
applying a stirring laser ($A=-30$) the energy per particle is $E(A=-30)\,=\,8.709$ (\hw). Now the difference in 
energies is $\Delta E\,=\,E(A=-30)-E(A=0)=7.375$ (\hw), which is equivalent to 
$\Delta E=7.375\,\hbar\omega_{ho}=1.216\times 10^{-31}$ Joule per particle; a value much smaller than that of FT. 
Second, the change in temperature from an initial value $T_0$ can be estimated from

\begin{equation}
\Delta T\,=\,T-T_0\,=\frac{N\Delta E}{C(T_c)},
\label{eq:deltaE}
\end{equation}

where $T_c$ is substituted into Eq.~(\ref{eq:specific-heat-ideal-Bose-gas}). It is recalled that $T_c$ 
for the ideal 2D BEC in a harmonic trap is given by \cite{Dalfovo:1999}

\begin{equation}
T_c\,=\,\frac{\hbar\omega_{ho}}{k_B} \sqrt{\frac{N}{\xi(2)}},
\label{eq:Tc}
\end{equation}

with $k_B$ Boltzmann's constant and $\xi(2)=1.6449$. For $N\sim 117$, $T_c=10.12$ nK. 
Considering $\omega_x=\omega_y=\omega_{ho}$ and that $T_0=0$, one gets $\Delta T=T=2.018$ nK. 
The condensate fraction is then estimated from

\begin{equation}
N\,=\,N_0\left(1\,-\,\frac{T^2}{T_c^2}\right),
\label{eq:condensate-fraction}
\end{equation}

and is equivalent to $N/N_0=0.96$. Therefore, our $T$ is indeed small. See, for example Neely \ea\ 
\cite{Neely:2010} and Onofrio \ea\ \cite{Onofrio:2000} where temperatures of $T=52$ nK and $T=10$ nK, 
respectively, were reported.

\hs There exist methods for beyond mean-field examinations of BEC dynamics in the group of L. Cederbaum, 
e.g. by B\u{r}ezinov\'{a}  \ea\ \cite{Brezinova:2012} who explored the expansion of a BEC in shallow 1D 
potentials using the TDGPE and the multi-configurational time-dependent Hartree for bosons (MCTDHB) \cite{Alon:2008} 
methods. It has been shown, that the onset of wave chaos in the GPE can be used as an indication for 
condensate depletion. The authors particularly focused on the case where the condensate depletion is 
relatively weak $\stackrel{<}{\sim} 5\%$. So far, it is known that as condensate depletion increases, 
the GPE becomes less valid as one faces a many-body problem beyond the GPE. However, B\u{r}ezinov\'{a} 
\ea\ made a comparison between the dynamics of the GPE and the MCTDHB and revealed that the mean-field 
effect of wave chaos --i.e., the buildup of random fluctuations-- corresponds to the many-body effect 
of condensate depletion. An important and surprising finding has been that there is good agreement 
between expectation values of observables obtained by GPE and MCTDHB, 
such as the width of the cloud and the kinetic energy. It has been further found, that the GPE can mimick 
excitations out of the condensate, and although the depletion lies outside the range of GPE applicability, 
one can monitor the onset of depletion by the onset of wave chaos within GPE, a fact that extends the range 
of GPE applicability. Further work involving beyond-the-GPE treatments has been presented, e.g., by Billam
\ea\ \cite{Billam:2012,Billam:2013} in which a second-order number-conserving numerical method
has been applied to solve the equations of motion involving a coupling between the condensate and
noncondensate. Their goal was to explore finite-temperature BEC dynamics and their method has been
successfully applied to the $\delta-$kicked rotor BEC.

\section{Summary and conclusions}\label{sec:conclusions}

\hs In summary, conditions have been obtained under which order and chaos appear in the dynamics of 
interacting trapped Bose gases. This work has specifically distinguished chaos in coordinate space 
from that in energy space. The chief result is that either quantum effects or trap anharmonicity is 
a generator of chaos in energy space. This conclusion has been reached through an artificial variation 
of the relative Planck's constant $\Gamma$ to values smaller or larger than 1 following 
Ref.~\cite{Kapulkin:2008}. A second important result is that chaos has been confirmed in the energy 
space of an excited trapped BEC. For severely reduced quantum effects ($\Gamma\ll 1$) in the presence 
of an external harmonic trap, no chaos is observed in either coordinate or energy space. Therefore, 
one way of suppressing chaos is by increasing the characteristic scale associated with the external 
trap with respect to the condensate size. Therefore, trap harmonicity in the absence of quantum effects 
is a generator of complete order in the physical observables. The presence of an external anharmonic 
trap and severely reduced quantum effects yields chaos in energy space, but not in coordinate space. 
The same happens in the presence of quantum effects ($\Gamma=1$) and trap harmonicity. Therefore, to 
obtain chaos in coordinate space, both quantum effects (with $\Gamma>1$) and trap anharmonicity (with 
$p_1=p_2>1$) must be present. This can also be inferred from Table~\ref{tab:tablogic}, which turns out 
to be similar to the logic-OR table. It is noted that, even if chaos exists in the energy space of a 
trapped BEC, it does not necessarily translate to chaos in coordinate space. Likewise, order in 
coordinate space does not imply order in energy space.

\begin{table}
\caption{Overview of conditions for order and chaos in coordinate (RMS radius 
$\sqrt{\langle r^2\rangle}$) and energy $(\langle E\rangle)$ space under the possibility of an 
artificial variation of the relative Planck's constant [$\Gamma=(\hbar_0/\hbar)^2$] following 
Ref.~\cite{Kapulkin:2008}. The system is a BEC in an external trap cut off by a hard-wall BP 
boundary. It is excited by a stirring laser (see main text). From left to right, the table lists 
conditions: QE?= quantum effects?, TA?= trap anharmonicity?, and the results for the presence or 
absence of chaos in both spaces. Answers are either yes (Y) or no (N).}
\begin{center}
\begin{tabular}{*{4}{|c}|} \hline
QE? & TA? & Chaos $\sqrt{\langle r^2\rangle}$? & Chaos $\langle E\rangle$?\\ \hline\hline
N & N & N & N \\ \hline
N & Y & N & Y \\ \hline
Y & N & N & Y \\ \hline
Y & Y & Y & Y \\ \hline
\end{tabular}
\end{center}
\label{tab:tablogic}
\end{table}

\hs Other results are as follows:

\begin{enumerate}

\item The non-periodic trajectories of $\dot{X}$ vs $X$ ($X$ being any physical quantity) supported by 
positive Lyapunov exponents are manifestations of chaos in the physical observables.

\item The frequency of oscillation of a property $X$ is primarily determined by the external trap. In the 
presence of a stirring blue-detuned laser, this frequency is not affected, whereas a stirring red-detuned 
laser changes the frequency. A dynamically changing effective trapping frequency is found to be a source 
of chaos in the BEC.

\item While the stirring laser is inside the trap, this situation could be viewed as an initial condition. 
If one considers measuring these systems after the removal of the stirrer, then one can think of different 
initializations according to whether there was a stirring BDLP or RDLP. Inspecting the post-stirring 
dynamics in Fig.~\ref{fig:ComponentsplotEvstG10B4T100Step5percentKseveralA}, one can infer that these 
systems are able to {\it remember} the kind of laser potential used to excite them. It turns out that 
the dynamics of the BEC is determined according to its {\it history} of excitations. As the trajectories 
that a chaotic system follows are sensitive to initial conditions \cite{Wimberger:2014,Brezinova:2011}, 
then this further confirms that our systems are indeed chaotic.

\end{enumerate}

\hs The usefulness of the present work is that: (1) it looks deeper into the chaotic dynamics of a BEC by 
looking at the dynamics of the energy components; (2) the ideas presented here can be used to gain further 
understanding of other analogous complex systems, such as the recently achieved photonic BEC \cite{Klaers:2010}; 
(3) it should motivate the exploration of chaos excited by other methods, such as an oscillating stirrer 
\cite{Fujimoto:2010,Fujimoto:2011,Raman:1999,Onofrio:2000} and a rotational one 
\cite{Caradoc:1999,Caradoc:2000,Raman:2001,Madison:2001}.

\section{Acknowledgments}
\hs The authors thank the Abdus Salam International Center for Theoretical Physics in Trieste, Italy for a 
hospitable stay during which part of this work was undertaken. ARS thanks the Max Planck Institute for Physics 
of Complex systems (MPIPKS) in Dresden Germany for a hospitable stay and for providing access to their excellent 
computing facilities on which most of the current simulations were performed. Stimulating and enlightening 
discussions with Soskin Stanislav (Lancaster University, UK) and Rajat Karnatak (MPIPKS Dresden, Germany) are 
gratefully acknowledged. This work was undertaken during sabbatical leave granted to the author Asaad R. Sakhel 
from Al-Balqa Applied University (BAU) during academic year 2014/2015. AB acknowledges financial support by the 
Ministry of Education, Science, and Technological Development of the Republic of Serbia under project ON171017.

\begin{appendix}

\section{Effective trapping frequency}\label{app:effective-trapping-frequency}

\hs The trapping frequencies $\omega_x$ and $\omega_y$ of the combined external+laser trap $\tilde{V}(x,y;t)$, 
Eq.~(\ref{eq:combined-potential}), are given by

\begin{equation}
\omega_q(x,y;t)\,=\,\sqrt{\frac{\partial^2 \tilde{V}(x,y;t)}{\partial q^2}},
\label{eq:overall-trapping-frequency}
\end{equation}

where $q\equiv(x,y)$ and $\omega_q$ is therefore a function of the coordinates. However, these equations 
are only valid near the minima of $\tilde{V}(x,y;t)$. Substituting Eq.~(\ref{eq:combined-potential}) into 
Eq.~(\ref{eq:overall-trapping-frequency}) yields

\begin{eqnarray}
&&\omega_x(x,y;t)\,=\,\left\{\frac{\sigma}{4}p_1(p_1-1)|x|^{p_1-2}\,-\,\right.\nonumber\\
&&\left.2\beta A(1\,-\,2\beta x^2)\,\exp[-\beta(x^2+(y-vt)^2)]\right\}^{1/2},\nonumber\\
\label{eq:omega_x.dependence_on_p}
\end{eqnarray}

and

\begin{eqnarray}
&&\omega_y(x,y;t)\,=\,\left\{\frac{\sigma}{4}p_2(p_2-1)|y|^{p_2-2}\,-\,\right.\nonumber\\
&&\left.2\beta A[1\,-\,2\beta(y-vt)^2]\exp[-\beta(x^2+(y-vt)^2)]\right\}^{1/2}.\nonumber\\
\label{eq:omega_y.dependence_on_p}
\end{eqnarray}

Hence, $\omega_x$ and $\omega_y$ are controlled by the overall shape of the combined trap and particularly 
the height or depth of the applied laser potential. Note that for $A<0$, $\omega_x$ and $\omega_y$ will 
increase with ``increasing" $A<0$. If $A>0$, the frequencies decrease.

\hs By setting $p_1=p_2=2$, one obtains for a harmonic trap

\begin{eqnarray}
&&\omega_x(x,y;t)\,=\,\nonumber \\
&&\left\{\frac{\sigma}{2}\,-\,2\beta A (1\,-\,2\beta x^2) \exp[-\beta(x^2\,+\,(y-vt)^2)]\right\}^{1/2},\nonumber\\
\label{eq:omega_y_at_p=2}
\end{eqnarray}

and

\begin{eqnarray}
&&\omega_y(x,y;t)\,=\,\nonumber\\
&&\left\{\frac{\sigma}{2}-2\beta A[1\,-\,2\beta(y-vt)^2]\exp[-\beta(x^2+(y-vt)^2)]\right\}^{1/2}.\nonumber\\
\label{eq:omega_y_at_p=2}
\end{eqnarray}

Note that in this case, the spatial variations of the combined trap ruling $\omega_q$ arise only from the 
laser potential because those due to the harmonic trap have been eliminated!

\subsection{Red-detuned Laser}

\hs At $t=0$, the only minimum in the combined harmonic trap is found at $x=y=0$ at the bottom of the RDLP 
well. Therefore

\begin{equation}
\omega_x(0,0;0)\,=\,\omega_y(0,0;0)\,=\,\left[\frac{\sigma}{2}\,-\,2\beta A\right]^{1/2}.
\label{eq:omegax000omegay000RDLP}
\end{equation}

At $t>0$, when the laser has moved only a little, such that there is still only one minimum in the combined 
trap, one gets at $x=0$ and $y=vt$ the same $\omega_x$ and $\omega_y$ as in Eq.~(\ref{eq:omegax000omegay000RDLP}), 
i.e.,

\begin{equation}
\omega_x(0,vt;t)\,=\,\omega_y(0,vt;t)\,=\,\left[\frac{\sigma}{2}\,-\,2\beta A\right]^{1/2}.
\label{eq:omegax0vttRDLP}
\end{equation}

As long as there is only one minimum (that of the RDLP), $\omega_x$ and $\omega_y$ will remain constant 
at all times $t$. However, when the red-detuned laser has moved far enough from the center of the harmonic 
trap, another minimum arises at $x=y=0$. Here, the frequencies become time-dependent with values given by

\begin{eqnarray}
&&\omega_x(0,0;t)\,=\,\left\{\frac{\sigma}{2}\,-\,2\beta A \exp[-\beta v^2 t^2]\right\}^{1/2},
\label{eq:omegax00tRDLP}
\end{eqnarray}

and

\begin{eqnarray}
&&\omega_y(0,0;t)\,=\,\left\{\frac{\sigma}{2}\,-\,2\beta A[1\,-\,2\beta v^2 t^2]\exp[-\beta v^2 t^2]\right\}^{1/2}.
\nonumber\\
\label{eq:omegay00tRDLP}
\end{eqnarray}

Hence the frequency of BEC-density oscillations inside the trap is subject to change with time, and this tends 
to be one source of chaos in these oscillations. Inside the reference frame of the RDLP, the BEC oscillates at 
a fixed frequency when $p_1=p_2=2$.

\hs For the anharmonic trap, say with $p_1=p_2=7$, at the minimum of the RDLP $x=y=0$, one obtains for $t=0$

\begin{equation}
\omega_x(0,0;0)\,=\,\omega_y(0,0;0)\,=\,\sqrt{-2\beta A},
\label{eq:omegax}
\end{equation}

and similarly for $t>0$

\begin{equation}
\omega_x(0,vt;t)\,=\,\omega_y(0,vt;t)\,=\,\sqrt{-2\beta A}.
\label{eq:omegay}
\end{equation}

When the RDLP has moved far away from $x=y=0$, there arises a minimum at the center of the anharmonic trap 
with trapping frequencies

\begin{eqnarray}
&&\omega_x(0,0;t)\,=\,\sqrt{-2\beta A}\,\exp\left(-\frac{1}{2}\beta v^2 t^2\right), \nonumber\\
&&\omega_y(0,0;t)\,=\,\sqrt{-2\beta A(1\,-\,2\beta v^2 t^2)}\,\exp\left(-\frac{1}{2}\beta v^2 t^2\right), \nonumber\\
\end{eqnarray}

and inside the RDLP

\begin{eqnarray}
&&\omega_y(0,vt;t)\,=\,\left[10.5\sigma |vt|^5 -2\beta A\right]^{1/2}, \nonumber\\
&&\omega_x(0,vt;t)\,=\,\sqrt{-2\beta A}. \nonumber\\
\label{eq:omega_y}
\end{eqnarray}

\subsection{Blue-detuned laser}

\hs In this case for a harmonic trap at $t=0$, we have a maximum at $x=y=0$, and there exists a minimum along 
a circular region around the barrier of $\tilde{V}(x,y;t)$. Assuming that this circle of minima has a radius 
$r_0$, then $x_0^2+y_0^2=r_0^2$ and $t=0$ yield

\begin{equation}
\omega_x(x_0,y_0;0)=\left\{\frac{\sigma}{2}-2\beta A(1-2\beta x_0^2)\exp[-\beta r_0^2]\right\}^{1/2},
\end{equation}

and
\begin{equation}
\omega_y(x_0,y_0;0)=\left\{\frac{\sigma}{2}-2\beta A(1-2\beta y_0^2)\exp[-\beta r_0^2]\right\}^{1/2}.
\end{equation}

When the BDLP moves, the previous circle of minima will vanish, and once the BDLP is far enough from the center 
of the harmonic trap the minimum at $x=y=0$ reappears. There is still a second minimum between the BDLP and 
the BP when viewed along the $y-$direction. If this minimum is located at $y_0$ and time $t$, then it is 
possible that this minimum with trapping frequency

\begin{eqnarray}
&&\omega_y(0,y_0;t)\,=\,\nonumber\\
&&\left\{\frac{\sigma}{2}\,-\,2\beta A[1\,-\,2\beta(y_0-vt)^2]\exp[-\beta(y_0-vt)^2]\right\},
\nonumber\\
\label{eq:omegay0y0tBDLP}
\end{eqnarray}

could provide some trapping at $y_0$ along $y$. However, in the $x-$direction Eq.~(\ref{eq:overall-trapping-frequency}) 
no longer applies for this case. The latter extremum is a saddle point with negative curvature in the $x-$direction and 
positive curvature in the $y-$direction.

\end{appendix}

\bibliography{chaos}

\begin{thebibliography}{94}

\bibitem{Kodba:2005}
{Stane Kodba, Matja$\check{z}$ Perc and Marko Marhl}, European Journal of
  Physics \textbf{26}, 205 (2005)

\bibitem{Filho:2000}
{Victo S. Filho, A. Gammal, T. Frederico, and Lauro Tomio}, Phys. Rev. A
  \textbf{62}, 033605 (2000)

\bibitem{Gertjerenken:2010}
{Bettina Gertjerenken, Stephan Arlinghaus, Niklas Teichmann, and Christoph
  Weiss}, Phys. Rev. A \textbf{82}, 023620 (2010)

\bibitem{Muruganandam:2002}
{Paulsamy Muruganandam and Sadhan K. Adhikari}, Phys. Rev. A \textbf{65},
  043608 (2002)

\bibitem{Xiong:2010}
{Hongwei Xiong and Baio Wu}, Phys. Rev. A \textbf{82}, 053634 (2010)

\bibitem{Katz:2010}
{Nadav Katz and Odeg Adam}, New J. Phys. \textbf{12}, 073020 (2010)

\bibitem{Chong:2004}
{G. Chong, W. Hai, and Q. Xie}, Chaos \textbf{14}, 217 (2004)

\bibitem{Brezinova:2011}
{Iva Brezinov{\'a}, Lee A. Collins, Katharina Ludwig, Barry I. Schneider, and
  Joachim Burgd{\"o}rfer}, Phys. Rev. A \textbf{83}, 043611 (2011)

\bibitem{Tomsovic:1991}
{Steven Tomsovic and Eric J. Heller}, Phys. Rev. Lett. \textbf{67}, 664 (1991)

\bibitem{Diver:2014}
{M. Diver, G. R. M. Robb, and G.-L. Oppo}, Phys. Rev. A \textbf{89}, 033602
  (2014)

\bibitem{Jaouadi:2010}
{A. Jaouadi, N. Gaaloul, B. Viaris de Lesegno, M. Telmini, L. Pruvost, and E.
  Charron}, Phys. Rev. A \textbf{82}, 023613 (2010)

\bibitem{Zhu:2009}
{Qianquan Zhu, Wenhua Hai, and Shiguang Rong}, Phys. Rev. E \textbf{80}, 016203
  (2009)

\bibitem{Martin:2009}
{J. Martin, B. Georgeot, and D. L. Shepelyansky}, Phys. Rev. E \textbf{79},
  066205 (2009)

\bibitem{Horsely:2014}
{Eric Horsely, Stewart Koppell, and L. E. Reichl}, Phys. Rev. E \textbf{89},
  012917 (2014)

\bibitem{Coullet:2001}
{P. Coullet and N. Vandenberghe}, Phys. Rev. E \textbf{64}, 025202(R) (2001)

\bibitem{Kapulkin:2008}
{Arie Kapulkin and Arjendu K. Pattanayak}, Phys. Rev. Lett. \textbf{101},
  074101 (2008)

\bibitem{Zhang:2011}
{Jing Zhang, Yu-xi Liu, Wei-Min Zhang, Lian-Ao Wu, Re-Bing Wu, and Tzyh-Jong
  Tarn}, Phys. Rev. B \textbf{84}, 214304 (2011)

\bibitem{Liu:1997}
{W. Vincent Liu and William C. Schieve}, Phys. Rev. Lett. \textbf{78}, 3278
  (1997)

\bibitem{Brambila:2013}
{D. S. Brambila and A. Fratalocchi}, Sci. Rep. \textbf{3}, 1 (2013)

\bibitem{Brandstaeter:1983}
{A. Brandst\"ater, J. Swift, Harry L. Swinney, A. Wolf, T. Doyne Farmer, and P.
  J. Crutchfield}, Phys. Rev. Lett. \textbf{51}, 1442 (1983)

\bibitem{Gardiner:2002}
{S. A. Gardiner}, J. Mod. Opt. \textbf{49}, 1971 (2002)

\bibitem{Cheng:2010}
{Jing Cheng}, Phys. Rev. A \textbf{81}, 023619 (2010)

\bibitem{Hai:2008}
{Wenhua Hai, Shiguang Rong, and Qianquan Zhu}, Phys. Rev. E \textbf{78}, 066214
  (2008)

\bibitem{Brezinova:2012}
{Iva Brezinov{\'a}, Axel U. J. Lode, Alexej I. Streltsov, Ofir E. Alon, Lorenz
  S. Cederbaum, and Joachim Burgd{\"o}rfer}, Phys. Rev. A \textbf{86}, 013630
  (2012)

\bibitem{Levi:2003}
{B. L$\acute{e}$vi, B. Georgeot, and D. L. Shepelyansky}, Phys. Rev. E
  \textbf{67}, 046220 (2003)

\bibitem{Georgeot:2000}
{B. Georgeot and D. L. Shepelyansky}, Phys. Rev. E \textbf{62}, 3504 (2000)

\bibitem{Weitenberg:2011}
{Christof Weitenberg, Stefan Kuhr, Klaus M{\o}lmer, and Jacob F. Sherson},
  Phys. Rev. A \textbf{84}, 032322 (2011)

\bibitem{Pachos:2003}
{Jiannis K. Pachos and Peter L. Knight}, Phys. Rev. Lett. \textbf{91}, 107902
  (2003)

\bibitem{Milner:2001}
{V. Milner, J. L. Hanssen, W. C. Campbell, and M. G. Raizen}, Phys. Rev. Lett.
  \textbf{86}, 1514 (2001)

\bibitem{Zhang:2004}
{Chuanwei Zhang, Jie Liu, Mark G. Raizen,and Qian Niu}, Phys. Rev. Lett.
  \textbf{92}, 054101 (2004)

\bibitem{Wimberger:2014}
{Sandro Wimberger}, \emph{{Nonlinear {D}ynamics and {Q}uantum {C}haos}}
  ({Springer Int. Pub.}, {Switzerland}, 2014)

\bibitem{Chacon:2008}
{R. Chac{\'o}n, D. Bote, and R. Carretero-Gonz{\'a}lez}, Phys. Rev. E
  \textbf{78}, 036215 (2008)

\bibitem{Castin:1997}
{Y. Castin and R. Dum}, Phys. Rev. Lett. \textbf{79}, 3553 (1997)

\bibitem{Gardiner:2000}
{S. A. Gardiner, D. Jaksch, R. Dum, J. I. Cirac, and P. Zoller}, Phys. Rev. A
  \textbf{62}, 023612 (2000)

\bibitem{Billam:2012}
{T. P. Billam and S. A. Gardiner}, New J. Phys. \textbf{14}, 013038 (2012)

\bibitem{Billam:2013}
{T. P. Billam, P. Mason and S. A. Gardiner}, Phys. Rev. A \textbf{87}, 033628
  (2013)

\bibitem{Liu:2006}
{Jie Liu, Chuanwei Zhang, Mark G. Raizen, and Qian Niu}, Phys. Rev. A
  \textbf{73}, 013601 (2006)

\bibitem{Fallani:2004}
{L. Fallani, L. De Sarlo, J. E. Lye,M. Modugno, R. Saers, C. Fort, and M.
  Inguscio}, Phys. Rev. Lett. \textbf{93}, 140406 (2004)

\bibitem{Ferris:2008}
{Andrew J. Ferris, Matthew J. Davis, Reece W. Geursen, P. Blair Blakie, and
  Andrew C. Wilson}, Phys. Rev. A \textbf{77}, 012712 (2008)

\bibitem{Garrett:2011}
{Michael C. Garrett, Adrian Ratnapala, Eikbert D. van Ooijen, Christopher J.
  Vale, Kristian Weegink, Sebastian K. Schnelle, Otto Vainio, Norman R.
  Heckenberg, Halina Rubinsztein-Dunlop, and Matthew J. Davis}, Phys. Rev. A
  \textbf{83}, 013630 (2011)

\bibitem{Stiessberger:2000}
J.S. Stie{\ss}ßberger, W.~Zwerger, Phys. Rev. A \textbf{62}, 061601(R) (2000)

\bibitem{Fujimoto:2010}
K.~Fujimoto, M.~Tsubota, Phys. Rev. A \textbf{82}, 043611 (2010)

\bibitem{Fujimoto:2011}
K.~Fujimoto, M.~Tsubota, Phys. Rev. A \textbf{83}, 053609 (2011)

\bibitem{Jackson:2000}
{B. Jackson, J. F. McCann, and C. S. Adams}, Phys. Rev. A \textbf{61},
  051603(R) (2000)

\bibitem{Radouani:2004}
{Abdelaziz Radouani}, Phys. Rev. A \textbf{70}, 013602 (2004)

\bibitem{Caradoc:1999}
{B. M. Caradoc-Davies, R. J. Ballagh, and K. Burnett}, Phys. Rev. Lett.
  \textbf{83}, 895 (1999)

\bibitem{Caradoc:2000}
{B. M. Caradoc-Davies, R. J. Ballagh, and P. B. Blakie}, Phys. Rev. A
  \textbf{62}, 011602(R) (2000)

\bibitem{Neely:2010}
{T. W. Neely, E. C. Samson, A. S. Bradley, M. J. Davis, and B. P. Anderson},
  Phys. Rev. Lett. \textbf{104}, 160401 (2010)

\bibitem{Engels:2007}
P.~Engels, C.~Atherton, Phys. Rev. Lett. \textbf{99}, 160405 (2007)

\bibitem{Onofrio:2000}
{R. Onofrio, C. Raman, J. M. Vogels, J. R. Abo-Shaeer, A. P. Chikkatur, and W.
  Ketterle}, Phys. Rev. Lett. \textbf{85}, 2228 (2000)

\bibitem{Madison:2000}
{K. W. Madison, F. Chevy, W. Wohlleben, and J. Dalibard}, Phys. Rev. Lett.
  \textbf{84}, 806 (2000)

\bibitem{Raman:2001}
{C. Raman, J. R. Abo-Shaeer, J. M. Vogels, K. Xu, and W. Ketterle}, Phys. Rev.
  Lett. \textbf{87}, 210402 (2001)

\bibitem{Raman:1999}
{C. Raman, M. K\"{o}hl, R. Onofrio, D. S. Durfee, C. E. Kuklewicz, Z.
  Hadzibabic, and W. Ketterle}, Phys. Rev. Lett. \textbf{83}, 2502 (1999)

\bibitem{Madison:2001}
{K. W. Madison, F. Chevy, V. Bretin, and J. Dalibard}, Phys. Rev. Lett.
  \textbf{86}, 4443 (2001)

\bibitem{Horng:2009}
{T.-L. Horng, S.-C. Gou, T.-C. Lin, G. A. El, A. P. Itin, and A. M.
  Kamchatnov}, Phys. Rev. A \textbf{79}, 053619 (2009)

\bibitem{Proukakis:2006}
{N. P. Proukakis, J. Schmiedmayer, and H. T. C. Stoof}, Phys. Rev. A
  \textbf{73}, 053603 (2006)

\bibitem{Diener:2002}
{Roberto B. Diener, Biao Wu, Mark G. Raizen, and Qian Niu}, Phys. Rev. Lett.
  \textbf{89}, 070401 (2002)

\bibitem{Aioi:2011}
{Tomohiko Aioi, Tsuyoshi Kadokura, Tetsuo Kishimoto, and Hiroki Saito}, Phys.
  Rev. X \textbf{1}, 021003 (2011)

\bibitem{Uncu:2008}
{Haydar Uncu, Devrim Tarhan, Ersan Demiralp, {\"O}zgur E.
  M{\"u}stecaplio$\tilde{g}$lu}, Las. Phys. \textbf{18}, 331 (2008)

\bibitem{Carpentier:2008}
{A. V. Carpentier, J. Belmonte-Beitia, H. Michinel, M. I. Rodas-Verde}, J. Mod.
  Opt. \textbf{55}, 2819 (2008)

\bibitem{Hammes:2002}
{M. Hammes, D. Rychtarik, H.-C. N{\"a}gerl, and R. Grimm}, Phys. Rev. A
  \textbf{66}, 051401(R) (2002)

\bibitem{Scherer:2007}
{David R. Scherer, Chad N. Weiler, Tyler W. Neely, and Brian P. Anderson},
  Phys. Rev. Lett. \textbf{98}, 110402 (2007)

\bibitem{Tuchendler:2008}
{C. Tuchendler, A. M. Lance, A. Browaeys, Y. R. P. Sortais, and P. Grangier},
  Phys. Rev. A \textbf{78}, 033425 (2008)

\bibitem{Stamper-Kurn:1998}
{D. M. Stamper-Kurn, H.-J. Miesner, A. P. Chikkatur, S. Inouye, J. Stenger, and
  W. Ketterle}, Phys. Rev. Lett. \textbf{81}, 2194 (1998)

\bibitem{Comparat:2006}
{D. Comparat, A. Fioretti, G. Stern, E. Dimova, B. Laburthe Tolra, and P.
  Pillet}, Phys. Rev. A \textbf{73}, 043410 (2006)

\bibitem{Jacob:2011}
{D. Jacob, E. Mimoun, L. De Sarlo, M. Weitz, J. Dalibard, and F. Gerbier}, New
  J. Phys. \textbf{13}, 065022 (2011)

\bibitem{Gustavson:2002}
{T. L. Gustavson, A. P. Chikkatur, A. E. Leanhardt, A. G{\"o}rlitz, S. Gupta,
  D. E. Pritchard, and W. Ketterle}, Phys. Rev. Lett. \textbf{88}, 020401
  (2001)

\bibitem{Barrett:2001}
{M. D. Barrett, J. A. Sauer, and M. S. Chapman}, Phys. Rev. Lett. \textbf{87},
  010404 (2001)

\bibitem{Schulz:2007}
{M. Schulz, H. Crepaz, F. Schmidt-Kaler, J. Eschner and R. Blatt}, J. Mod. Opt.
  \textbf{54}, 1619 (2007)

\bibitem{Parker:2003}
{N. G. Parker, N. P. Proukakis, M. Leadbeater, and C. S. Adams}, Phys. Rev.
  Lett. \textbf{90}, 220401 (2003)

\bibitem{Parker:2010}
{N. G. Parker, N. P. Proukakis, and C. S. Adams}, Phys. Rev. A \textbf{81},
  033606 (2010)

\bibitem{Proukakis:2004}
{N. P. Proukakis, N. G. Parker, C. F. Barenghi, and C. S. Adams}, Phys. Rev.
  Lett. \textbf{93}, 130408 (2004)

\bibitem{Law:2000}
{C. K. Law, C. M. Chan, P. T. Leung, and M.-C. Chu}, Phys. Rev. Lett.
  \textbf{85}, 1598 (2000)

\bibitem{Sakhel:2011}
{Roger R. Sakhel, Asaad R. Sakhel, and Humam B. Ghassib}, Phys. Rev. A
  \textbf{84}, 033634 (2011)

\bibitem{Sakhel:2013}
{Roger R. Sakhel, Asaad R. Sakhel, Humam B. Ghassib}, J. Low. Temp. Phys.
  \textbf{173}, 177 (2013)

\bibitem{Muruganandam:2009}
{P. Muruganandam and S. K. Adhikari}, Computer Physics Communications
  \textbf{180}, 1888 (2009)

\bibitem{Vudragovic:2012}
{D. Vudragovi\'c, I. Vidanovi\'c, A. Bala\v z, P. Muruganandam, and S. K.
  Adhikari}, Computer Physics Communications \textbf{183}, 2021 (2012)

\bibitem{Ruostekoski:2001}
{J. Ruostekoski, B. Kneer, W. P. Schleich, and G. Rempe}, Phys. Rev. A
  \textbf{63}, 043613 (2001)

\bibitem{Bongs:1999}
{Bongs, K. and Burger, S. and Birkl, G. and Sengstock, K. and Ertmer, W. and
  Rz{\c a}{\. z}ewski, K. and Sanpera, A. and Lewenstein, M.}, Phys. Rev. Lett.
  \textbf{83}, 3577 (1999)

\bibitem{Deng:2012}
{Yan Deng, Wenhua Hai, Gengbiao Lu and Shiguang Rong}, J. Phys. B: At. Mol.
  Opt. Phys. \textbf{45}, 135301 (2012)

\bibitem{Dutta:2015}
{Dutta, Shovan and Mueller, Erich J.}, Phys. Rev. A \textbf{91}, 013601 (2015)

\bibitem{Dalfovo:1999}
{F. Dalfovo, S. Giorgini, L. P. Pitaevskii, and S. Stringari}, Rev. Mod. Phys.
  \textbf{71}, 463 (1999)

\bibitem{Pethick:2002}
{C. J. Pethick and H. Smith}, \emph{{Bose-Einstein Condensation in Dilute
  Gases}} ({Cambridge University Press}, {Cambridge}, 2002)

\bibitem{Blakie:2008}
{P. B. Blakie, A. S. Bradley, M. J. Davis, R. J. Ballagh, and C. W. Gardiner},
  Adv. Phys. \textbf{57}, 363 (2008)

\bibitem{Donley:2001}
{Elizabeth A. Donley, Neil R. Claussen, Simon L. Cornish, Jacob L. Roberts,
  Eric A. Cornell, and Carl E. Wieman}, Nature \textbf{412}, 295 (2001)

\bibitem{gpe-integrability}
{Viewed from a spatial point of view then, the GPE is integrable as its
  behavior is predictable in the latter space.}

\bibitem{Quasiperiodicity-chaos-transition}
{Indeed, in Ref.\cite{Diver:2014} it was argued that the appearance and
  disappearance of chaos is due to the transition from quasiperiodic behavior
  to frequency locking and vice versa. When the RDLP leaves the box potential
  at $t>5$, the oscillations of $E_{zp}$ and $E_{flow}$ remain irregular since
  the BEC is in a highly excited state and frequency locking is absent. }

\bibitem{Mateos:1998}
{Jos$\acute{e}$ L. Mateos and Jorge V. Jos$\acute{e}$}, Physica A \textbf{257},
  434 (1998)

\bibitem{Rossler:1976}
{O. E. R{\"o}ssler}, Phys. Lett. A \textbf{A57}, 397 (1976)

\bibitem{Amaral:2006}
{Gleison F. V. Amaral, Christophe Letellier, and Luis Antonio Aguirre}, Chaos
  \textbf{16}, 013115 (2006)

\bibitem{Sasaki:2010}
K.~Sasaki, N.~Suzuki, H.~Saito, Phys. Rev. Lett. \textbf{104}, 150404 (2010)

\bibitem{Schreck:2001}
{F. Schreck, L. Khaykovich, K. L. Corwin, G. Ferrari, T. Bourdel, J.
  Cubizolles, and C. Salomon}, Phys. Rev. Lett. \textbf{87}, 080403 (2001)

\bibitem{Alon:2008}
{Ofir E. Alon, Alexej I. Streltsov, and Lorenz S. Cederbaum}, Phys. Rev. A
  \textbf{77}, 033613 (2008)

\bibitem{Klaers:2010}
{Jan Klaers, Julian Schmitt, Frank Verwinger, and Martin Weitz}, Nature
  \textbf{468}, 545 (2010)

\end{thebibliography}

\end{document}